\documentclass[conference]{IEEEtran}
\ifCLASSINFOpdf
\else
\fi
\usepackage{fancyhdr}

\usepackage[sort,nocompress]{cite}
\usepackage[hyphens]{url}
\usepackage{graphicx}
\usepackage{textcomp}
\usepackage{xcolor}
\usepackage[bookmarks=true,breaklinks=true,letterpaper=true,colorlinks,linkcolor=black,citecolor=blue,urlcolor=black]{hyperref}

\usepackage[normalem]{ulem}

\usepackage{soul}
\usepackage{algorithm}
\usepackage{xspace}
\usepackage[font=footnotesize,labelfont=bf]{caption}
\usepackage{mathtools}
\usepackage{amsfonts}
\usepackage{amssymb}
\usepackage{pifont}
\usepackage{color}
\usepackage{algpseudocode}
\usepackage{multirow}
\usepackage{textgreek}
\usepackage{subfig}
\usepackage{soul}

\newcommand{\fig}[1]{Figure~\ref{#1}}
\newcommand{\sect}[1]{Section~\ref{#1}}
\newcommand{\tab}[1]{Table~\ref{#1}}
\newcommand{\algo}[1]{Algorithm~\ref{#1}}

\newcommand{\proposed}[0]{Hera\xspace}

\newcommand{\ncf}[0]{\texttt{NCF}\xspace}
\newcommand{\din}[0]{\texttt{DIN}\xspace}
\newcommand{\dien}[0]{\texttt{DIEN}\xspace}
\newcommand{\wnd}[0]{\texttt{WnD}\xspace}

\newcommand{\dlrm}[1]{\texttt{DLRM({#1})}\xspace}
\newcommand{\model}[1]{\emph{Model$_{#1}$}\xspace}
\newcommand{\cacheway}[1]{\emph{CacheWay$_{#1}$}\xspace}
\newcommand{\membw}[1]{\emph{MemBW$_{#1}$}\xspace}

\newcommand{\coaff}[1]{\emph{CoAff$_{#1}$}\xspace}
\newcommand{\loadtime}[1]{\emph{T$_{#1}$}\xspace}

\newcommand{\deeprecsys}[0]{\texttt{DeepRecSys}\xspace}
\newcommand{\random}[0]{\texttt{Random}\xspace}
\newcommand{\randomplus}[0]{\texttt{Hera(Random)}\xspace}
\newcommand{\hera}[0]{\texttt{Hera}\xspace}

\algnewcommand{\LineComment}[1]{\State \(\triangleright\) #1}

\newcommand\blfootnote[1]{%
\begingroup
\renewcommand\thefootnote{}\footnote{#1}%
\addtocounter{footnote}{-1}%
\endgroup
}

\def\BibTeX{{\rm B\kern-.05em{\sc i\kern-.025em b}\kern-.08em
    T\kern-.1667em\lower.7ex\hbox{E}\kern-.125emX}}

\renewcommand{\baselinestretch}{.999}
\title{\LARGE Hera: A Heterogeneity-Aware Multi-Tenant Inference Server \\ for Personalized Recommendations}

\begin{document}

\author{

\IEEEauthorblockN{
Yujeong Choi\hspace{2em}John Kim\hspace{2em}Minsoo Rhu}
\IEEEauthorblockA{
School of Electrical Engineering\\
KAIST\\
\texttt{\{yjchoi0606, jjk12, mrhu\}@kaist.ac.kr}\\
}
}

\maketitle
\pagestyle{plain}

\begin{abstract}

While providing low latency is a fundamental requirement in deploying
recommendation services, achieving high resource utility is also
crucial in cost-effectively maintaining the datacenter.  Co-locating multiple
workers of a model is an effective way to maximize query-level parallelism
and server throughput, but the interference caused by concurrent workers at
shared resources can prevent server queries from meeting its SLA.  Hera
utilizes the heterogeneous memory requirement of multi-tenant recommendation
models to intelligently determine a productive set of co-located models and its
resource allocation, providing fast response time while achieving high
throughput.  We show that Hera achieves an average $37.3\%$ improvement in
effective machine utilization, enabling $26\%$ reduction in required servers,
					significantly improving upon the baseline recommedation inference
					server.

\end{abstract}

\IEEEpeerreviewmaketitle
\blfootnote{
Preprint.\\
}
\section{Introduction}

Deep neural network (DNN) based personalized recommendation models play a vital
role in today's consumer facing internet services (e.g., e-commerce, news feed,
		Ads).  Facebook, for instance, reports that recommendation models account
for more than $75\%$ of all the machine learning (ML) inference cycles in their
datacenters~\cite{gupta2020architectural}.  
A major challenge facing this emerging ML workload is the need to effectively
balance low latency and high throughput.  More concretely, unlike training
scenarios where throughput is the primary figure-of-merit, ensuring low latency
responsiveness is a fundamental requirement for inference services,
especially for these user-facing recommendation models.  Nonetheless, achieving
high server utility and system throughput is still vital for hyperscalers as
cost-effectively maintaining the consolidated datacenters directly translates
into low total cost of ownership (TCO).

	Given this landscape, ``co-locating'' multiple \emph{workers} from a
	single or multiple recommendation models is an effective solution to
	improve system throughput. As the inference server is constantly being
	fed with numerous service queries, the scheduler can utilize such
	\emph{query-level parallelism} to have multiple inference queries be
	\emph{concurrently} processed using these multiple workers.
	Recommendations are typically deployed using CPUs because of
	their high availability at datacenters as well as their
	latency-optimized design.  A {\bf key challenge} in co-locating
	recommendation models over a multi-core CPU is determining which models
	to co-locate together, how many workers per each model to deploy, and
	how to gracefully handle the interference between co-located workers at
	shared resources, i.e., caches and the memory system.
	Latency-critical ML tasks operate with strict service level agreement
	(SLA) goals on tail latency, so even a small amount of disturbance at
	shared resources can cause deteriorating effects on ``latency-bounded''
	throughput (i.e., the \emph{number of queries processed per second that
	meets SLA targets}, aka QPS).

To this end, an important motivation and contribution of this work is a
detailed characterization on the effect of co-locating multiple recommendation
model workers on tail latency as well as latency-bounded throughput.
	We make several observations unique to multi-tenant
recommendation inference.  Conventional convolutional and recurrent neural
networks (CNNs and RNNs) are primarily based on highly regular DNN 
algorithms. These ``dense'' DNN models enjoy high QPS improvements by
utilizing query-level parallelism to scale up the number of multi-tenant
workers~\cite{tripti:2019:trtis}, a property this paper henceforth refers to as
\emph{worker scalability}.  However, DNN-based recommendations employ
``sparse'' \emph{embedding layers} in addition to dense DNNs, exhibiting a
highly irregular memory access pattern over a large embedding table.  Depending
on which application domain the recommendation model is being deployed (e.g.,
ranking, filtering, $\ldots$), the configuration of different models can vary
significantly in terms of its 1) embedding table size, 2) the number of
embedding table lookups per each table, and 3) the depth/width of the dense DNN
layers.
All these factors determine the memory capacity and bandwidth demands of a
model, affecting its worker scalability (\sect{sect:characterization}). For
example, recommendations with a modest model size generally exhibit a
  compute-limited, cache-sensitive behavior with high worker scalability.  On
the other hand, models with high memory (capacity and/or bandwidth)
  requirements are substantially limited with their worker scalability,
  rendering co-location a suboptimal, or worse, an impossible design point.  As
  a result, blindly choosing the recommendation models to co-locate, without
  accounting for each model's worker scalability, leads to aggravated tail
  latency and QPS, leaving significant performance left on the table.

In this paper, we present \emph{Hera}, a ``{\bf He}terogeneous Memory {\bf
	R}equirement {\bf A}ware Co-location Algorithm'' for multi-tenant
	recommendation inference. The innovation of \proposed lies in its
	ability to accurately estimate \emph{co-location affinity} among a given pair
	of recommendation models.  Models with high (or low) co-location affinity are
	defined as those that can (or fail to) sustain high
	QPS while sharing the compute/memory resources with each
	other. Our {\bf key observation} is that 
	memory capacity and/or bandwidth
	limited recommendations with low worker
	scalability generally exhibit high co-location affinity with 
	models with high worker scalability. This is because
	memory-limited models fail to fully utilize on-chip cores and
	its local caches, allowing compute-limited/cache-sensitive 
	models to leverage such opportunity to spawn more workers while causing less
	disturbance to tail latency.

	Based on such key observation, we design \proposed with two key
		components: 1) a ``cluster-level'' model selection unit
			(\sect{sect:hera_model_selection}) and 2) a ``node-level'' shared
			resource management unit (\sect{sect:hera_resrc_manager}). At the
			cluster level, \proposed utilizes an analytical model that systematically
			evaluates co-location affinity among any given pair of recommendation
			models. \proposed's model selection unit then utilizes this information
			to determine \emph{what models} are most appropriate to be co-located
			together across all the nodes within the cluster. Once the models to
			co-locate are determined, \proposed's node-level resource management unit
			examines \emph{how many workers} as well as \emph{how much shared cache
				capacity} each model should be allocated within each node that
        effectively utilize query-level parallelism for high sustained QPS.
        Overall, \proposed achieves an average $37.3\%$ improvement in
        effective machine utilization, which enables a $26\%$ reduction in the
        number of required inference servers, significantly improving
        state-of-the-art.

\begin{figure}[t!] \centering
\includegraphics[width=0.38\textwidth]{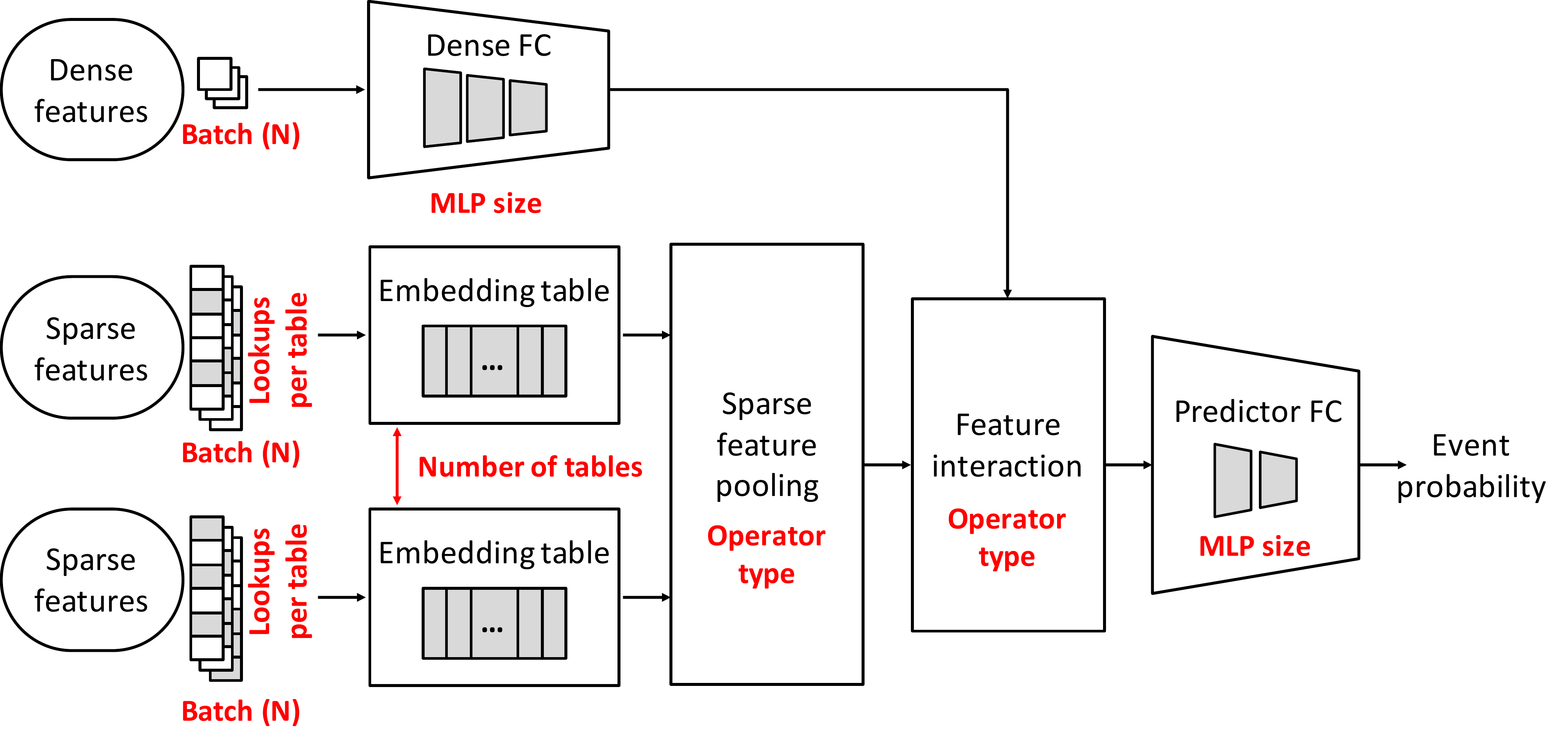}
\vspace{-.5em}
\caption{
Model architecture of DNN-based recommendations
}
\vspace{-1.5em}
\label{fig:recsys_topology}
\end{figure}

\section{Background}
\label{sect:background}

\subsection{Neural Recommendation Models}
\label{sect:dnn}

Recommendation models aim to find out contents/items to recommend to a user
based on prior interactions as well as user's preference.  
A well-known challenge with content
				 recommendations is that users interact only with a subset of available
				 contents and items. Take YouTube as an example where any given user
				 only watches a tiny subset of available video clips. Consequently,
				 state-of-the-art DNN-based recommendation models combine both
				 \emph{dense} and \emph{sparse} features for high accuracy. Here, dense
				 features represent continuous inputs (e.g., user's age) whereas sparse
				 features represent categorical inputs (e.g., a collection of movies a
						 user has previously watched).  \fig{fig:recsys_topology} shows a
				 high-level overview of DNN-based recommendation models which we detail
				 below.

{\bf Model architecture overview.}
The dense features in recommendations are processed with a stack of dense
(bottom) DNN layers (e.g., convolutions, recurrent, and MLPs).  Categorical
features on the other hand are encoded as multi-hot vectors where a ``1''
represent a positive interaction among the available contents/items. As the
number of positive interaction among all possible items are extremely small
(i.e., a small number of ``1''s within the multi-hot vector), the multi-hot
vectors are transformed into real-valued, dense vectors (called
		\emph{embeddings}) by an \emph{embedding layer}. Specifically, an array of
embedding vectors are stored contiguously as a table, and a sparse index ID
(designating the location of ``1''s within the multi-hot vector) is used to
read out a unique row from this table. Because the multi-hot vector is
extremely sparse, reading out the embedding vectors (corresponding to each
		sparse indices) from the table is equivalent to a sparse vector
\emph{gather} operation.  The embedding vectors gathered from a given embedding
table are  reduced down into a single vector using element-wise additions.  In
general, embedding vector gather operations exhibit a highly memory-limited
behavior as they are highly sparse and irregular memory accesses.  Another
distinguishing aspect of embedding layers is their high memory capacity
demands: the embedding tables can contain several millions of entries, amounting to tens to hundreds of GBs of
memory usage.

As there are multiple embedding tables, multiple \emph{reduced} embeddings
are generated by the embedding layer, the result of which goes  through
a feature interaction stage. One popular mechanism for feature interactions is a
dot-product operation~\cite{naumov2019deep} between all input vectors (i.e., implemented as a batched
		GEMM).  The feature interaction output is concatenated with the output
of the bottom DNN layer, which is subsequently processed by the top DNN layers to calculate
an event probability (e.g., the click-through-rates in advertisement banners) for
recommending contents/items.

\begin{figure}[t!] \centering
\includegraphics[width=0.30\textwidth]{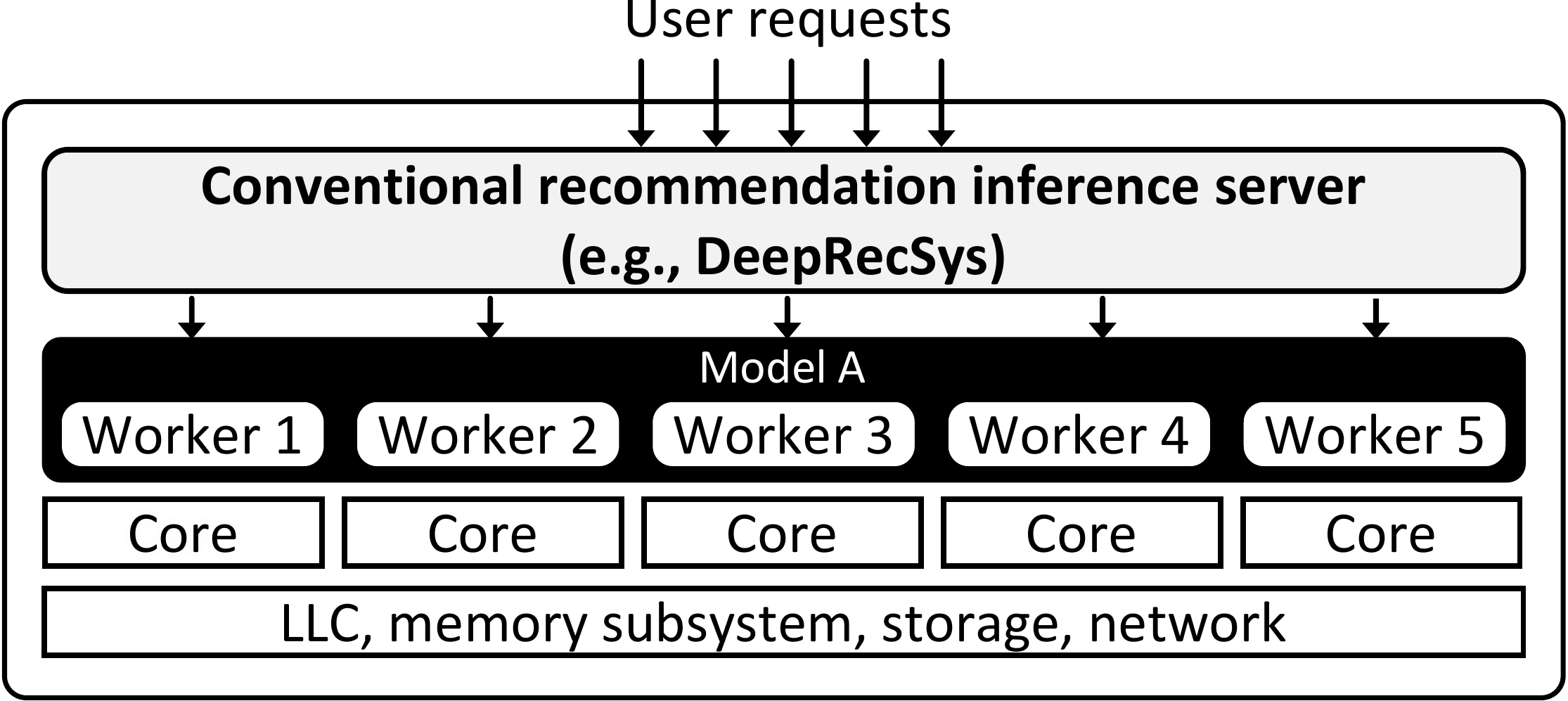}
\caption{
A CPU-based multi-tenant recommendation inference server.
}
\vspace{-1.3em}
\label{fig:deep_rec_sys_overview}
\end{figure}

	\begin{table*}[t!]
\centering
\caption{Key architectural configurations of state-of-the-art neural recommendation models.}
\vspace{-0.5em}
\scriptsize
\begin{tabular}{|c|c|c|c|c|c|c|c|c|c|c|}
\hline
\multirow{2}{*}{\textbf{Model}} & \multirow{2}{*}{\textbf{Domain}} & \multicolumn{3}{c|}{\textbf{FC}}                             & \multicolumn{5}{c|}{\textbf{Embeddings}}                                                       & \multicolumn{1}{l|}{\multirow{2}{*}{\textbf{SLA (ms)}}} \\ \cline{3-10}
                                &                                  & \textbf{Dense-FC} & \textbf{Predict-FC} & \textbf{Size (MB)} & \textbf{Tables} & \textbf{Lookup} & \textbf{Dimension} & \textbf{Size (GB)} & \textbf{Pooling} & \multicolumn{1}{l|}{}                                   \\ \hline
                                DLRM (A)                        & Social media                     & 128-64-64         & 256-64-1            & 0.2                & 8               & 80              & 64                 & 2                  & Sum              & 100                                                     \\ \hline
                                DLRM (B)                        & Social media                     & 256 -128-64       & 128-64-1            & 0.5                & 40              & 120             & 64                 & 25                 & Sum              & 400                                                     \\ \hline
                                DLRM (C)                        & Social media                     & 2560-1024-256-32  & 512-256-1           & 12                 & 10              & 20              & 32                 & 2.5                & Sum              & 100                                                     \\ \hline
                                DLRM (D)                        & Social media                     & 256-256-256       & 256-64-1            & 0.2                & 8               & 80              & 256                & 8                  & Sum              & 100                                                     \\ \hline
                                NCF                             & Movies                           & -                 & 256-256-128         & 0.6                & 4               & 1               & 64                 & 0.1                & Concat           & 5                                                       \\ \hline
                                DIEN                            & E-commerse                       & -                 & 200-80-2            & 0.2                & 43              & 1               & 32                 & 3.9                & Attention+RNN    & 35                                                      \\ \hline
                                DIN                             & E-commerse                       & -                 & 200-80-2            & 0.2                & 4               & 3               & 32                 & 2.7                & Attention+FC     & 100                                                     \\ \hline
                                Wide\&Deep                      & Play store                       & -                 & 1024-512-256        & 8                  & 27              & 1               & 32                 & 3.5                & Concat           & 25                                                      \\ \hline
\end{tabular}
\vspace{-1em}
\label{tab:model_config}
\end{table*}

{\bf State-of-the-art DNN-based recommendation models.} While
\fig{fig:recsys_topology} broadly captures the high-level architecture of
DNN-based recommendations, state-of-the-art model architectures deployed in
industry settings exhibit notable differences in terms of their key design
parameters (colored in \emph{red} in \fig{fig:recsys_topology}).
\tab{tab:model_config} summarizes our studied, industry-scale DNN-based
recommendation models published from Google, Facebook, and
Alibaba~\cite{wide_and_deep,gupta2020architectural,din,dien,ncf}. We further detail our evaluation methodology later in
\sect{sect:methodology}.

\subsection{Inference Serving Architectures}
\label{sect:serving_arch}

{\bf Why CPUs for recommendation inference?} 
As the complexity of ML algorithms increases, GPUs or dedicated ASICs are
gaining momentum in servicing
recommendations~\cite{park2018deep,deng2021low,kalamkar2020optimizing}.
Nonetheless,  CPUs are still popular deployment options for several
hyperscalars~\cite{hazelwood2018applied,gupta2020architectural,centaur,deeprecsys,deng2021low,gupta2021recpipe}
because the abundance of CPUs in today's datacenters make them appealing from a
TCO perspective, particularly during the off-peak portions of the diurnal cycle
where they remain idle.  In fact, key industrial organizations are constantly
developing system-level solutions that that better optimize CPU-based
recommendation inference~\cite{jizhi,hazelwood2018applied,deng2021low}.  Given
their importance in ML inference, this paper focuses on CPU-based inference
servers tailored for recommendations.

{\bf Multi-tenant inference server architecture.} While providing fast response
to end-users is vital for inference, achieving high server utility is
also crucial for cost-effectively maintaining the consolidated datacenters.
Co-locating multiple \emph{workers} of an ML model 
on a single machine is an effective solution to
improve server utility and throughput at the cost of aggravated latency.
In our baseline CPU-based, multi-tenant inference
server, a single worker (implemented using Caffe2 worker) is allocated with a dedicated CPU core and its
local caches, multiples of which share the last level cache (LLC) and the
memory subsystem (\fig{fig:deep_rec_sys_overview}).  
Because the inference server is consistently being requested
with multiple service queries, having multiple workers helps leverage
\emph{query-level parallelism} to simultaneously execute multiple inference
services, improving throughput.
As inference is a highly latency-sensitive operation, existing ML frameworks
are designed with an \emph{in-memory} processing model assuming 
the entire working set of a given
worker process is all captured inside DRAM (i.e., paging data in and out of disk
		swap space is a non-option).  Therefore, having multiple workers be
co-located within a single machine requires the CPU memory capacity to be large
enough to fully accommodate the aggregate memory usage of \emph{all} the concurrent
workers~\cite{tf_serving,trtis,deeprecsys}.

\section{Related Work}
\label{sect:related}

Improving server cost-efficiency via multi-tenancy has been studied extensively
in prior literature~\cite{bubbleup,paragon,heracles,kasture2014ubik,
  jiang2010combining,jiang2010complexity}.  As co-located multi-tenant tasks
  contend for shared resources, prior work has focused on how to minimize
  interference and performance unpredictability. A common approach is to
  co-locate a user-facing, latency-critical task with best-effort workloads
  (e.g., batch jobs), prioritizing the latency-critical task with higher QoS to
  meet SLA. While effective in guaranteeing QoS for latency-critical tasks,
  such solution suffers from low server utility in terms of the number of
  latency-critical tasks scheduled per servers. Consequently, recent
  work~\cite{parties,clite,twig} explored QoS-aware resource management
  techniques that can accommodate \emph{multiple} latency-critical tasks
  within a single machine. Unlike these prior work focusing on generic,
  latency-critical cloud services (e.g., Memcached, Sphinx, MongoDB, $\ldots$),
  our work focuses on ML-based recommendation inference servers, presenting our
  unique, application-aware \proposed architecture. More importantly, \proposed
  develops a novel analytical model that quantifies co-location affinity among
  a given pair of recommendation models for intelligently selecting models to
  co-locate.

More relevant to \proposed is recent work by Gupta et
al.~\cite{deeprecsys,gupta2020architectural}, which conducts a workload
characterization on the effect of co-locating multiple workers from a single,
                 \emph{homogeneous} recommendation model
                 (\fig{fig:deep_rec_sys_overview}).  As we detail in
                 \sect{sect:effect_multitenant}, co-locating workers from a
                 single model severely limits the worker scalability for memory
                 capacity or bandwidth limited recommendations.  To the best of
                 our knowledge, \proposed is the first to quantitatively
                 evaluate the effect of co-locating workers from both
                 homogeneous as well as \emph{heterogeneous} recommendation
                 models, developing a cluster-wide heterogeneous model
                 selection algorithm as well as a node-level QoS-aware resource
                 partitioning algorithm for recommendation inference servers.
                 While not directly related to recommendation inference, Choi
                 et al.~\cite{prema} and Ghodrati et al.~\cite{planaria}
                 studied an NPU-based multi-tenant inference for
                 CNNs/RNNs/Attentions using temporal~\cite{prema} and
                 spatial~\cite{planaria} multi-tasking, respectively.  There is
                 also several recent work proposing near-data processing  for
                 accelerating
                 recommendations~\cite{tensordimm,recnmp,tensorcasting,trim,recssd}.
                 In general, the key contributions of \proposed is orthogonal
                 to these prior work.

	\begin{table}[t!]
  \centering
  \caption{CPU server node configuration.}
\scriptsize
\vspace{-0.5em}
\begin{tabular}{|c|c|}
\hline
\textbf{Component}        & \textbf{Specification}                  \\ \hline
CPU model                 &  \ \ Intel(R) Xeon(R) Gold $6242$ CPU \ \ \\ \hline
Frequency                 & $2.80$ GHz                                 \\ \hline
Physical cores per socket & $16$                                      \\ \hline
Sockets                   & $2$                                       \\ \hline
SIMD                      & AVX-$512$                                 \\ \hline
Shared L3 cache size      & $22$ MB (11-ways)                          \\ \hline
DRAM capacity             & $192$ GB per socket                        \\ \hline
DRAM speed                & $2666$ MT/s                                \\ \hline
DRAM bandwidth            & $128$ GB/s                                 \\ \hline
Operating system          & CentOS Linux 7                          \\ \hline
Network bandwidth         & $10$ Gbps                                  \\ \hline
\end{tabular}
\vspace{-1.5em}
  \label{tab:server_config}
\end{table}

\section{Methodology}
\label{sect:methodology}

{\bf Hardware platform.}  We utilize a multi-node cluster containing one master
node and five compute nodes, allowing a total of ten inference servers to be
deployed (\tab{tab:server_config}).  The intra-node resource partitioning and
isolation are done using Linux's \emph{cpuset cgroups} for allocating specific
core IDs to a given model worker, and Intel's \emph{Cache Allocation
	Technology} (CAT) for LLC partitioning.

{\bf Software architecture.}
\proposed's runtime manager is implemented using
Facebook's open-sourced DeepRecInfra~\cite{deeprecsys}, a software
framework for designing ML inference
servers for at-scale neural  recommendation models. 

{\bf Query arrival rates.} Prior work~\cite{deeprecsys} reports that the query
arrival rates of recommendation services in a production datacenter follows a
Poisson distribution.  Similarly, MLPerf's cloud inference 
suite~\cite{mlperf} also employs a Poisson distribution in its inference query
traffic generator.  As such, our evaluation utilizes DeepRecInfra's inference
query traffic generator which issues requests to the inference server
based on a Poisson distribution, the query arrival rate of which is configured
as appropriate per our evaluation goals (detailed in \sect{sect:results}).

{\bf Query working set size.} The size of queries for recommendation inference
decides the number of items to be ranked for a given user, which determines
request batch size.
Prior work~\cite{deeprecsys} observes that the working set sizes 
for recommendation queries follow a unique 
distribution with a heavy tail effect. We utilize DeepRecInfra
to properly reflect such distribution,
having the
batch size of an inference range from $1$-$1024$~\cite{deeprecsys}.

{\bf Benchmarks.} We utilize the open-sourced recommendation models
provided with DeepRecInfra~\cite{deeprecsys} to 
construct eight industry-representative model architectures. 
The SLA target of each model is properly chosen
in accordance to prior work~\cite{deeprecsys,wide_and_deep,din,dien}
(\tab{tab:model_config}).  As discussed in {\sect{sect:dnn}}, recent
literature frequently points to large embedding
tables~\cite{aibox,recssd,yi2018factorized,lui2020understanding} with wide
embedding vectors ($32$ to $1024$
dimensions)~\cite{hestness2019beyond,tensordimm,sambanova:blog:2020} as it
helps improve algorithmic performance.  Two of our model architectures accomodate
such latest algorithmic developments: 1) {\dlrm{B}}
that models similar scale of memory capacity requirement to MLPerf's
\texttt{DLRM} (sized with tens of GBs of embedding tables), and  2) \dlrm{D} modeling embeddings with wide
vectors.

\begin{figure}[t!] \centering
\vspace{-0.5em}
\includegraphics[width=0.4\textwidth]{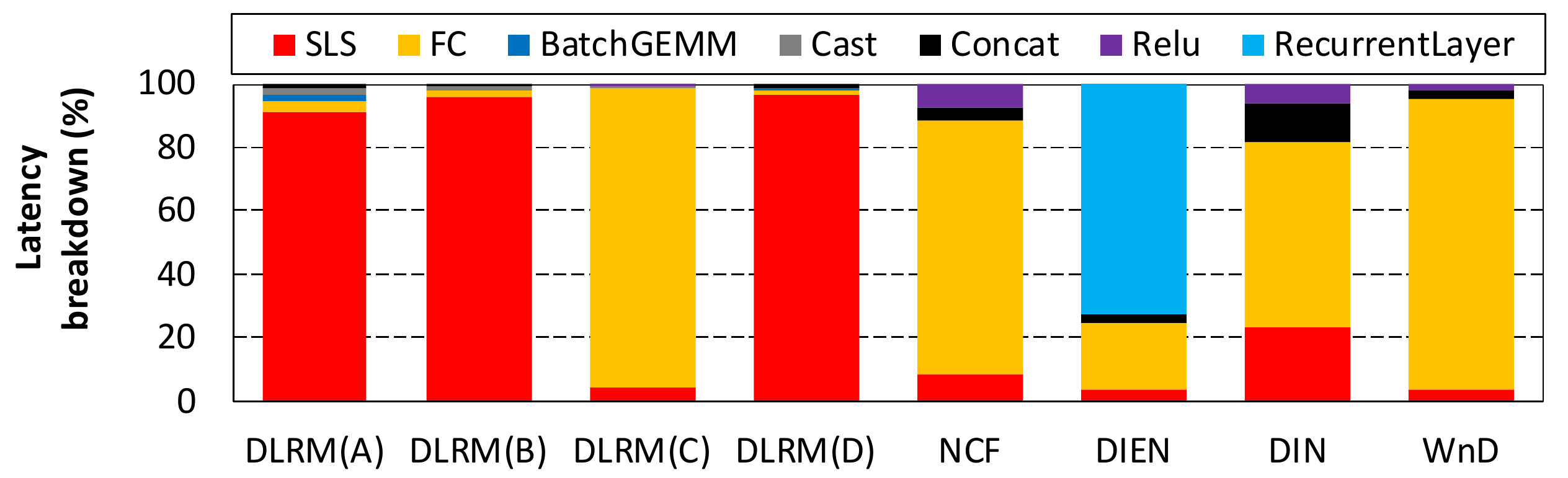}
\vspace{-0.3em}
\caption{
A single worker's inference time \emph{without} co-location broken into key 
operators of Caffe2, assuming a batch size of $220$ (mean value of our studied query size distribution,\sect{sect:methodology}).
Embedding layers are implemented using \texttt{{\bf S}parse{\bf L}engths{\bf S}um} (SLS) in Caffe2. FC and BatchGEMM refers to fully-connected and batched GEMM.
}
\vspace{-0.5em}
\label{fig:latency_breakdown_single_worker}
\end{figure}

\begin{figure}[t!] \centering
\vspace{-0.5em}
\includegraphics[width=0.48\textwidth]{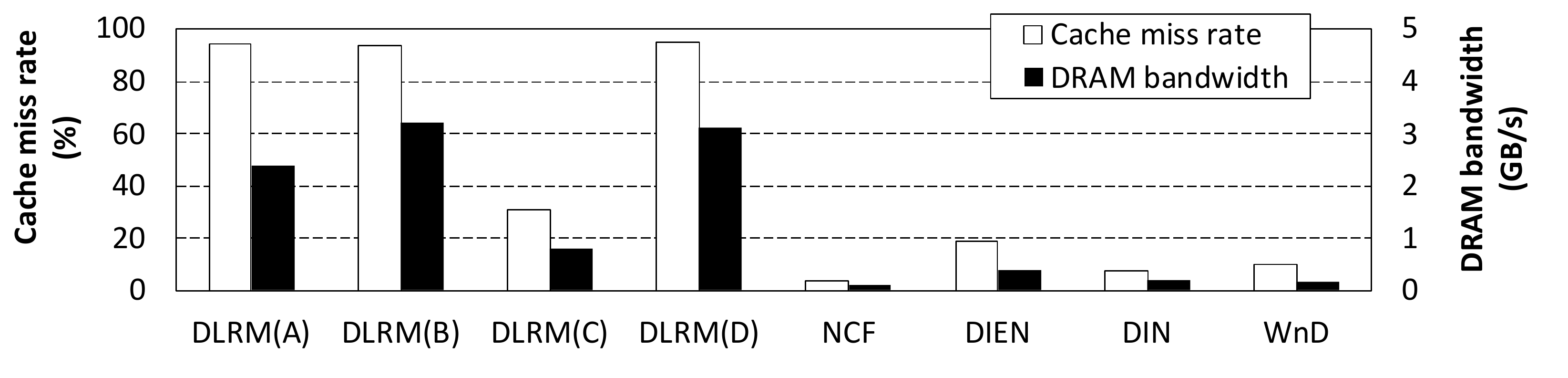}
\caption{
Effect of single worker inference on on-chip cache miss rate (left) and off-chip DRAM bandwidth utility (right). 
}
\vspace{-1.5em}
\label{fig:caching_dram_single_worker}
\end{figure}

\section{Workload Characterization}
\label{sect:characterization}

\subsection{Analysis on ``Single'' Model Worker}
\label{sect:single_worker}

We start by characterizing a single worker's inference behavior \emph{without}
co-location, breaking down latency per major compute operators.  As shown in
\fig{fig:latency_breakdown_single_worker}, the apparent operator diversity
leads to different performance bottlenecks. For example, models such as
\dlrm{A,B,D} are significantly bottlenecked on the memory intensive embedding
layers, experiencing high cache miss rate and high memory bandwidth usage
(\fig{fig:caching_dram_single_worker}). Because embedding vector gather
operations are conducted over a large embedding table, its memory access stream
are extremely sparse and irregular with low data locality.  Consequently,
    models with a large number of embedding tables and embedding lookups
    (\dlrm{A,B}) or wide embedding vectors (\dlrm{D}) show much higher memory
    intensity as the majority of execution time is spent gathering embedding
    vectors.  Such property is in stark contrast to \dlrm{C}, \ncf, \dien,
    \din, and \wnd, which spend significant amount of time on the
    computationally intensive FC and/or recurrent layers. Thanks to the
    (relatively) higher compute intensity and smaller working set, these
    compute-intensive models exhibit better caching efficiency and lower memory
    bandwidth consumption.

\begin{figure}[t!] 
\centering
\subfloat[]{
\includegraphics[width=0.44\textwidth]{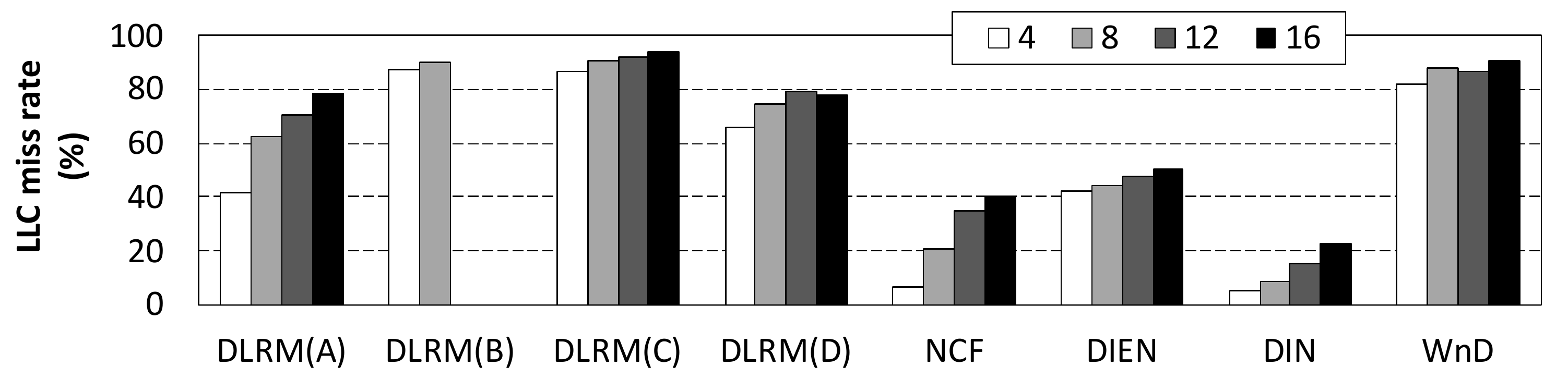}
	\label{fig:}
}
\vspace{-0em}
\subfloat[]{
	\includegraphics[width=0.44\textwidth]{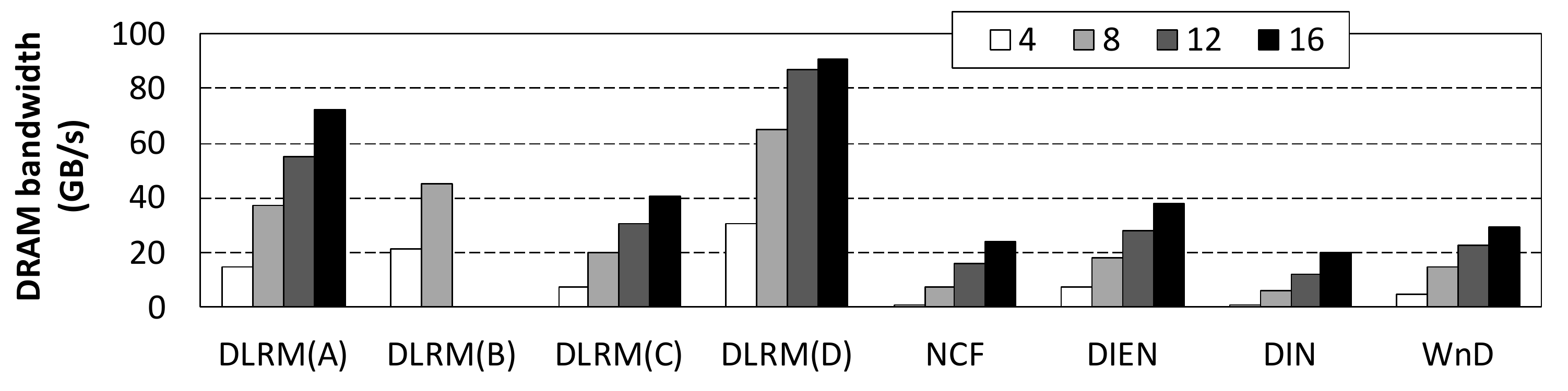}
	\label{fig:}
}
\vspace{-0.5em} 
\caption{ 
Effect of scaling up multi-tenant workers on (a) LLC miss rate and (b) memory
  bandwidth utilization. \dlrm{B} does not have bars under $12$/$16$ workers as
  it results in an out-of-memory error.  }   
\vspace{-0.75em}
\label{fig:sensitivity_memory_hierarchy}
\end{figure}

\subsection{Serving with ``Multi''-Tenant Workers}
\label{sect:effect_multitenant}

\begin{figure}[t!] \centering
\includegraphics[width=0.44\textwidth]{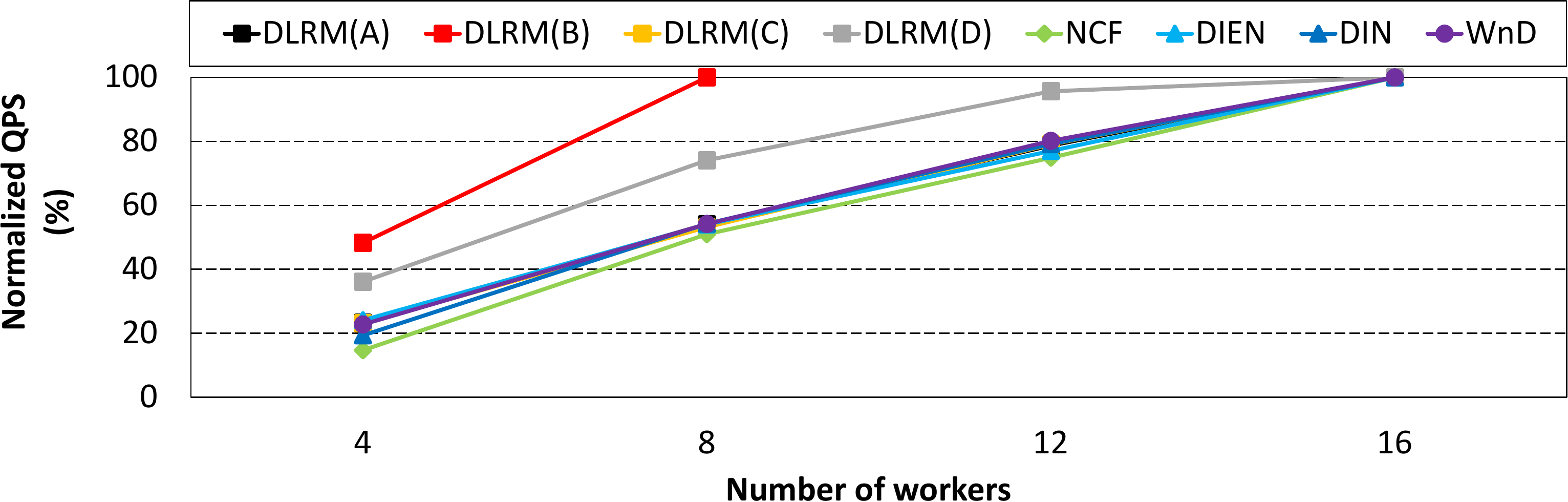}
\caption{
Latency-bounded throughput (QPS) as a function of the number of parallel
  workers. We show normalized QPS to $16$ workers to demonstrate each model's
  worker scalability.
}
\vspace{-1.5em}
\label{fig:worker_scalability}
\end{figure}

Building on top of the characterization on single worker inference, we now
study the efficacy of co-locating \emph{multiple} workers from a \emph{single}
recommendation model. We assume a single multi-core CPU machine is utilized for
servicing inference queries for the recommendation model but the number of
concurrently executing workers are scaled ``up'' (i.e., one worker per each
    core, \fig{fig:deep_rec_sys_overview}) to characterize its effect on shared
memory resources (\fig{fig:sensitivity_memory_hierarchy}) as well as
latency-bounded throughput, i.e., QPS (\fig{fig:worker_scalability}).  QPS is
measured by quantifying the maximum input load the concurrent workers can
process without violating SLA.  Specifically, we start from a low input query
arrival rate (i.e., queries arrived per second) and gradually inject higher
request rates until the observed ($95$th percentile) tail latency starts
violating the SLA target.  The \emph{max load} of a recommendation model
therefore is defined to quantify the maximum input query arrival rate its
workers are able to sustain without SLA violation.

As the number of workers are increased, we generally observe a gradual increase
in LLC miss rate with a corresponding increase in memory bandwidth usage. This
is expected as more workers proportionally demand larger compute and memory
usage.  However, there are noticeable differences in the way different models
react under a multi-tenant inference scenario.  First and foremost, memory
``capacity'' hungry models such as \dlrm{B}  is severely limited with its
ability to spawn a large number of concurrent workers as the aggregate memory
usage of a single worker alone amounts to $25$ GB (\tab{tab:model_config}).
Consequently, the inference server suffers from an \emph{out-of-memory} error
beyond $8$ workers (recall that ML inference servers employ a software stack
    implemented using an in-memory processing model, \sect{sect:serving_arch}),
       leaving an average $50\%$ of CPU cores and on-chip caches left idle.
       Such behavior is likely to be a significant concern to hyperscalers
       because recent literature frequently points to neural recommendation
       models having several tens to hundreds of GBs of memory
       usage~\cite{gupta2020architectural,lui2020understanding,yi2018factorized,zhao2020distributed}.
       As such, co-locating a large number of workers for these memory capacity
       limited models are challenging under current ML serving architectures
       (\fig{fig:worker_scalability}).  Second, embedding limited models with
       high memory ``bandwidth'' usage (\dlrm{A,D}) exhibit an almost linear
       increase in memory bandwidth utility with a large number of workers.
       This is because the LLC fails to capture the already meager data
       locality of embedding layers, frequently missing at the LLC and
       consuming high memory bandwidth.  The performance of \dlrm{D} in
       particular is completely bottlenecked on memory bandwidth, whose
       aggregated bandwidth usage saturates beyond $12$ workers
       (\fig{fig:sensitivity_memory_hierarchy}).  As a result, the QPS
       improvements in \dlrm{D} levels off around $12$ workers, only achieving
       a further $4\%$ throughput enhancements going from $12$ to $16$ workers.
       Therefore, scaling up the number of multi-tenant workers beyond $12$ for
       \dlrm{D} is highly sub-optimal from a throughput cost-efficiency
       perspective.  Lastly, the remaining five recommendation models with high
       compute intensity and a modest model size (\tab{tab:model_config}) leave
       plenty of memory bandwidth available even when the number of parallel
       workers are maximized. As shown in \fig{fig:worker_scalability}, such
       headroom in memory bandwidth allows these compute-intensive
       recommendation models to enjoy a scalable increase in QPS with a large
       number of parallel workers.

\begin{figure}[t!] \centering
\includegraphics[width=0.45\textwidth]{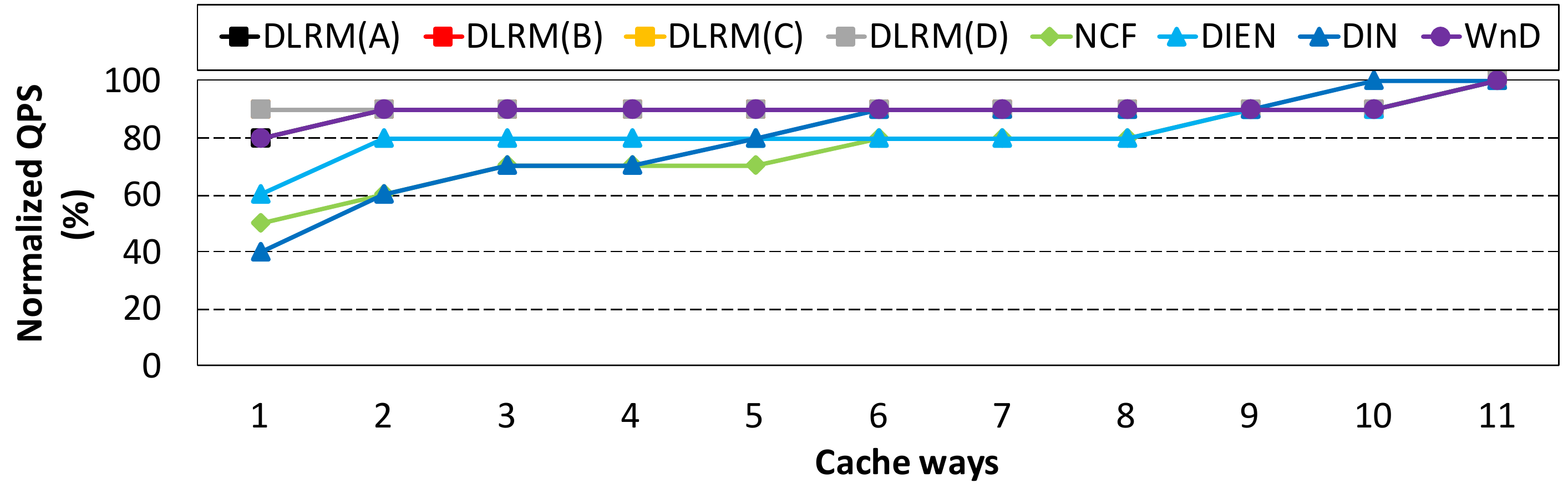}
\vspace{-0em}
\caption{
Changes in latency-bounded throughput (QPS) when limiting the number of LLC
  ways allocated to workers. Results are normalized to the right-most
  configuration where we allow the workers to fully utilize the entire (11
      ways) LLC.  Each experiment assumes maximally possible workers are
  spawned for execution (8 for DLRM (B) and 16 for other models).  Note that
  Intel's CAT prevents the allocation of \emph{zero} LLC ways to any given
  process (i.e., \emph{bypassing} the LLC is impossible), thus having at least
  a single LLC way allocated per each model is required.  
We observed similar trends when measuring sensitivity of QPS
on LLC over different number of workers, so we omit them for brevity.
}
\vspace{-1.5em}
\label{fig:cache_way_sensitivity}
\end{figure}

Overall, we conclude that recommendation models with high memory capacity
and/or bandwidth demands are severely limited with its \emph{worker
  scalability}, i.e., the ability to leverage query-level parallelism in ML
  inference servers to scalably increase QPS via multi-tenant workers. In
  contrast, models that are relatively more compute intensive with high cache
  sensitivity exhibit much better worker scalability.  Such high worker
  scalability however is only guaranteed when each worker is provided with a
  ``large enough'' LLC capacity to sufficiently capture locality, as we further
  discuss below.

\subsection{Sensitivity to LLC Capacity}
\label{sect:cache_sensitivity}

To analyze the sensitivity of each model's worker scalability to LLC size, we
utilize Intel's Cache Allocation Technology (CAT)~\cite{intel_cat} to limit the
number of LLC ways allocated to the multi-tenant workers.
\fig{fig:cache_way_sensitivity} plots the sustained QPS of each model
architecture (y-axis) as we gradually limit the number of ways allocated to the
executing workers (from right to left on the x-axis). Several important
observations can be made from this experiment.  First, the memory-limited
\dlrm{A,B,D} shows high robustness to available LLC capacity.  For instance,
  \dlrm{D} is able to achieve $90\%$ of maximum QPS despite having only a
  single LLC way allocated.  These memory-limited models spend significant
  fraction of their execution time on highly irregular embedding gather
  operations with low locality (\fig{fig:latency_breakdown_single_worker}).
  Therefore, leveraging memory parallelism rather than locality (i.e., memory
      bandwidth rather than  LLC capacity) is more crucial for these
  memory-limited models in achieving high sustained QPS.  Second, models with
  high compute-intensity (\ncf, \dien, \din, \wnd) exhibit high sensitivity to
  the LLC capacity, underscoring the importance of sufficiently provisioning
  shared cache space to the multi-tenant workers. Interestingly, many of these
  cache-sensitive workloads are able to sustain reasonably high QPS despite
  having allocated with a small LLC space. For instance, \dien is capable of
  achieving more than $80\%$ of maximum QPS while only being allocated with $2$
  out of the $11$ LLC ways (and similarly $2$ ways and $5$ ways for \wnd and
      \din, respectively).

\section{\proposed Architecture and Design} \label{sect:hera_arch}

\begin{figure}[t!] \centering
\includegraphics[width=0.48\textwidth]{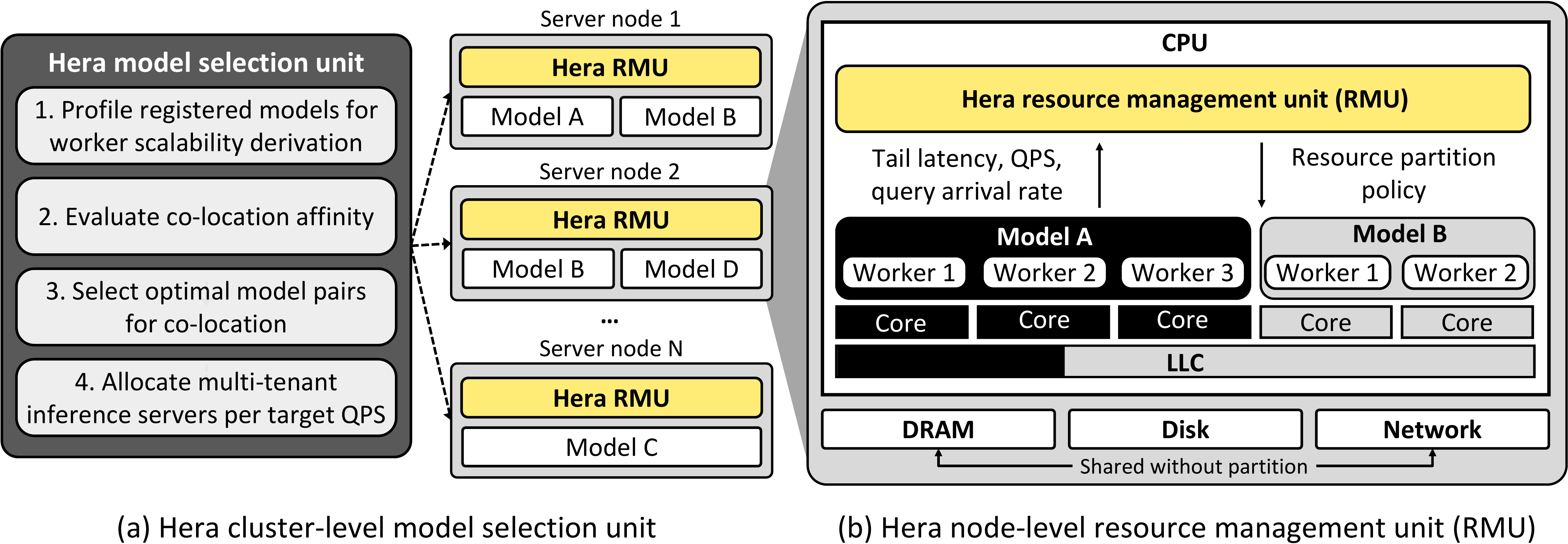}
\vspace{-0.2em}
\caption{
High-level overview of \proposed architecture. 
}
\vspace{-1.5em}
\label{fig:hera_overview}
\end{figure}

\subsection{Design Principles} 
\label{sect:overview} 

{\bf Overview.} \fig{fig:hera_overview} provides an overview of
  \proposed's two key components, 1) the \emph{cluster-level} model selection
  unit (\sect{sect:hera_model_selection}) and 2) the \emph{node-level} shared
  resource management unit (\sect{sect:hera_resrc_manager}). At the cluster
  level, \proposed employs our analytical model that utilizes the heterogeneous
  memory needs of recommendations for selecting the optimal pair of models to
  co-locate across all the server nodes.  Once the models to co-locate are
  determined, \proposed node-level resource management unit sets up the proper
  compute/memory resource allocation strategy for the multi-tenant workers,
  which is dynamically tuned for maximum efficiency by closely monitoring each
  model's tail latency, QPS, and query arrival rates.

\begin{figure}[t!] \centering
\includegraphics[width=0.49\textwidth]{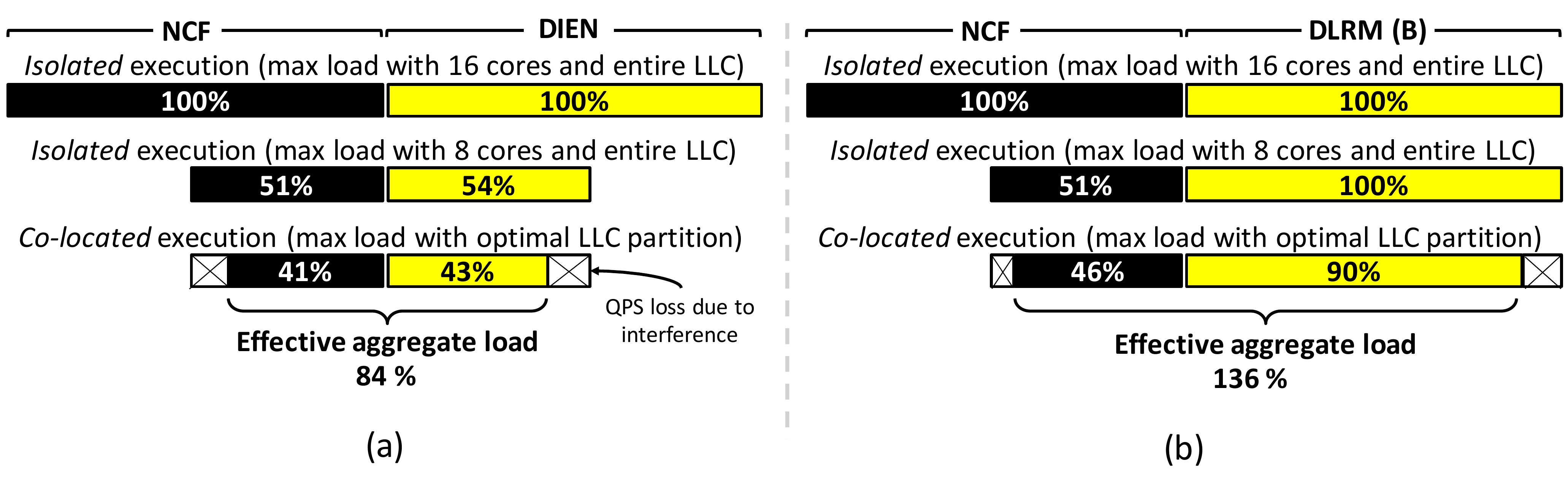}
\vspace{-1.5em}
\caption{
Effect of co-locating (a) (high, high) worker scalability models and (b) (high,
    low) worker scalability models. 	}
\vspace{-1.5em}
\label{fig:colocation_example}
\end{figure}

{\bf Key approach.} \sect{sect:effect_multitenant} revealed that models with
high memory capacity or bandwidth usage suffer from low worker scalability,
     leaving significant amount of CPU cores and shared LLC underutilized.  To
     address such inefficiency, \proposed aims to deploy workers from a pair of
     low and high worker scalability models, as their complementary compute and
     memory usage characteristic (i.e., memory-intensive vs.  compute-intensive
         execution for low vs. high worker scalability models respectively)
     helps maximize server utility while minimizing interference at shared
     resources. \fig{fig:colocation_example} provides examples that highlight
     the importance of \proposed's worker scalability aware, multi-tenant model
     selection algorithm, where the cache-sensitive \ncf is co-located with (a)
  another cache-sensitive \dien and (b) memory capacity limited \dlrm{B}.  As
  depicted, co-locating two cache-sensitive and high worker scalability models
  with similar resource requirements (\ncf and \dien) results in severe
  interference at the LLC, causing an average $20\%$ throughput loss compared
  to each model's isolated execution.  In contrast, consider the example in
  \fig{fig:colocation_example}(b) where workers with complementary
  compute/memory access patterns are co-located.  Because of \dlrm{B}'s memory
  capacity constraints, the model is not able to fully utilize the $16$ cores
  on-chip, rendering the remaining CPU cores and LLC left underutilized.
  Co-locating \ncf with \dlrm{B} therefore helps better utilize server
  resources thereby significantly improving aggregate throughput.  Of course,
  the interference between \dlrm{B} and \ncf cannot be eliminated completely,
  so both models experience some throughput loss compared to their respective
  isolated executions.  Nonetheless, the net benefit of enhanced server utility
  via intelligently co-locating low and high worker scalability models
  outweighs the deterioration in each model's throughput,  leading to a
  significant system-wide QPS improvement.

As such, we seek to address the key research challenge regarding how to
reliably estimate a given model's worker scalability anad utilize that
information to decide the optimal set of models to co-locate across the
cluster.  Once the models to co-locate are determined, another research
question remains regarding how to efficiently allocate shared resources among
multi-tenant workers that minimize interference and maximize QPS.  We first
elaborate on \proposed's cluster-level model selection algorithm in
\sect{sect:hera_model_selection} followed by a discussion of our node-lvel
resource management policy in \sect{sect:hera_resrc_manager}.

\subsection{Cluster-level Multi-tenant Model Selection Unit}
\label{sect:hera_model_selection}

{\bf Profiling-based worker scalability estimation.} Through our
characterization in \sect{sect:effect_multitenant}, we observe that a given
recommendation model's performance scalability (as a function of the number of
    concurrent workers) can be estimated reliably through profiling. \proposed
utilizes the slope of the performance scalability curve in
\fig{fig:worker_scalability} to make a binary decision on whether the subject
model has high worker scalability or not. For example, memory capacity limited
\dlrm{B} and memory bandwidth limited \dlrm{D} are categorized as those with
low worker scalability because employing a large number of workers is either
impossible (\dlrm{B}) or simply unproductive beyond a certain threshold from a
QPS perspective (\dlrm{D}).  The remaining recommendation models on the other
hand are categorized as having high worker scalability as they can fully
utilize the CPU cores and on-chip caches with sustained high QPS.  The profiled
result in \fig{fig:worker_scalability} only needs to be collected once for a
target server architecture and the derivation of whether a model has high
worker scalability or not is entirely done offline, having negligible impact on
\proposed's performance or memory usage, i.e., a static boolean variable per
each model designates its worker scalability.

{\bf Determining key sources of resource contention.} By utilizing the
recommendation model's (high/low) worker scalability information, \proposed
tries to co-locate a pair of (high, low) worker scalability models.  This helps
us significantly reduce the model selection search space as it tries to avoid
the unfruitful co-location of (high, high) worker scalability models.
Nonetheless, different recommendation models naturally exhibit different
compute and shared resource usage, so understanding a model's unique resource
requirement helps better estimate the magnitude of shared resource contention,
            thereby narrowing down the model selection search space even
            further.  Prior work~\cite{parties,clite,heracles} considers the
            shared LLC, memory, storage, and network bandwidth as one of the
            most important sources of resource contention under multi-tenancy.
            However, our key finding is that the multi-tenant workers for
            recommendation inference rarely compete for storage and network
            bandwidth. As previously discussed in \sect{sect:serving_arch},
            state-of-the-art ML frameworks for inference servers employ an
            in-memory processing model.  Consequently, once the ML inference
            server is bootstrapped for deployment (e.g., initializing inference
                server-client processes, provisioning each worker's memory
                allocation needs inside DRAM, $\ldots$), the multi-tenant
            workers have little to no interaction with the storage system at
            inference time. This is not surprising given the latency-critical
            nature of inference and its need for high performance
            predictability. When it comes to the network stack, the ML
            inference server receives client's service queries through the NIC
            over an HTTP/REST protocol~\cite{trtis,tf_serving}.  However, we
            observe less than an average $1.9$ Gbps of network bandwidth usage
            in all our evaluation, far less than the available system-wide
            network bandwidth (which is typically in the orders of several tens
                to hundreds of Gbps in at-scale datacenters).  Consequently,
            \proposed only considers the interference at shared LLC and memory
            bandwidth for evaluating both the candidate models for co-location
            and the efficient resource allocation mechanism for those models.
            Compared to previous, \emph{generic} QoS-aware resource
            partitioning mechanisms~\cite{heracles,parties} that consider the
            interference at all of cache/memory/storage/network, \proposed's
            \emph{application-awareness} helps our proposal more agiley adapt
            to the dynamics of inference query arrival patterns, significantly
            improving upon state-of-the-art (\sect{sect:eval_resrc_manager}).

\begin{algorithm}[t!]
\caption{Co-location Affinity}
\label{algo:colocation_affinity}
\begin{algorithmic}[1]
\scriptsize
\rmfamily
\LineComment{Step A: Derive co-location affinity at LLC}
	\State $\mathrm{CoAff_{LLC} = 0}$
	\For {$\mathrm{i = 1 \ \textbf{to} \ CacheWay_{max}}$}
	\State $\mathrm{CacheWay_{A} = i}$
	\State $\mathrm{CacheWay_{B} = CacheWay_{max} - CacheWay_{A}}$
	\State $\mathrm{CoAff_{temp} =} \newline
	\mathrm{(\dfrac{QPS[Model_{A}][CacheWay_{A}]}{QPS[Model_{A}][CacheWay_{max}]} +}
		\mathrm{\dfrac{QPS[Model_{B}][CacheWay_{B}]}{QPS[Model_{B}][CacheWay_{max}]})/2 }$
	\If {$\mathrm{CoAff_{temp} > CoAff_{LLC}}$}
		\State $\mathrm{CoAff_{LLC} = CoAff_{temp}}$
	\EndIf
\EndFor
\\
\LineComment{Step B: Derive co-location affinity at memory bandwidth}
	\State $\mathrm{CoAff_{DRAM} = \min(\dfrac{MemBW_{system}}{MemBW_{A} + MemBW_{B}}, 1)}$
\\
\LineComment{Step C: Estimation of system-level co-location affinity}
	\State $\mathrm{CoAff_{system} = \min(CoAff_{LLC}, CoAff_{DRAM})}$
\end{algorithmic}
\end{algorithm}

{\bf Identifying co-location affinity for model selection.} We develop an
analytical model that estimates \emph{co-location affinity} among a given pair
of models. \proposed's model selection unit utilizes co-location affinity to
determine the best set of models to co-locate, choosing those with high
co-location affinity, i.e., ones that can sustain high QPS while sharing the
LLC/memory.  Co-location affinity is derived by estimating how much QPS loss is
expected to the co-located \model{A} and \model{B} due to the interference at
the shared LLC and memory bandwidth -- the two most important sources of
resource contention under multi-tenant ML inference
(\algo{algo:colocation_affinity}).  Similar to our definition of worker
scalability, we employ a profiling-based approach to model the effect of shared
resource contention on QPS.  First, the profiled LLC sensitivity study in
\fig{fig:cache_way_sensitivity} is utilized to derive the expected QPS when
\model{A} and \model{B}  gets an equal partition of the CPU cores for worker
allocation. Each model is then given a partitioned slice of the shared LLC ways
(\cacheway{A} and \cacheway{B} in line $4$-$5$). The profiled QPS for each
model is then normalized to the QPS achieved when each model is given the
entire LLC for execution, the result of which is averaged over the two
co-located models to quantify the effect of LLC interference (line $6$).  By
examining all possible combination of LLC partitioning, we are able to derive
the most optimal LLC partitioning point that gives the highest aggregate QPS
(line $7$-$8$). As \proposed's resource management unit is capable of utilizing
such information for optimal LLC partitioning (detailed in
    \sect{sect:hera_resrc_manager}), we utilize this LLC partitioning point as
the reference to quantify co-location affinity at LLC.  As such, the closer the
\coaff{LLC} value is to $1$, the more likely the co-located models are less
interfering with each other, thus having high
co-location affinity at LLC.

\begin{figure}[t!] \centering
\vspace{-1.5em}
\subfloat[]{
\includegraphics[width=0.2\textwidth]{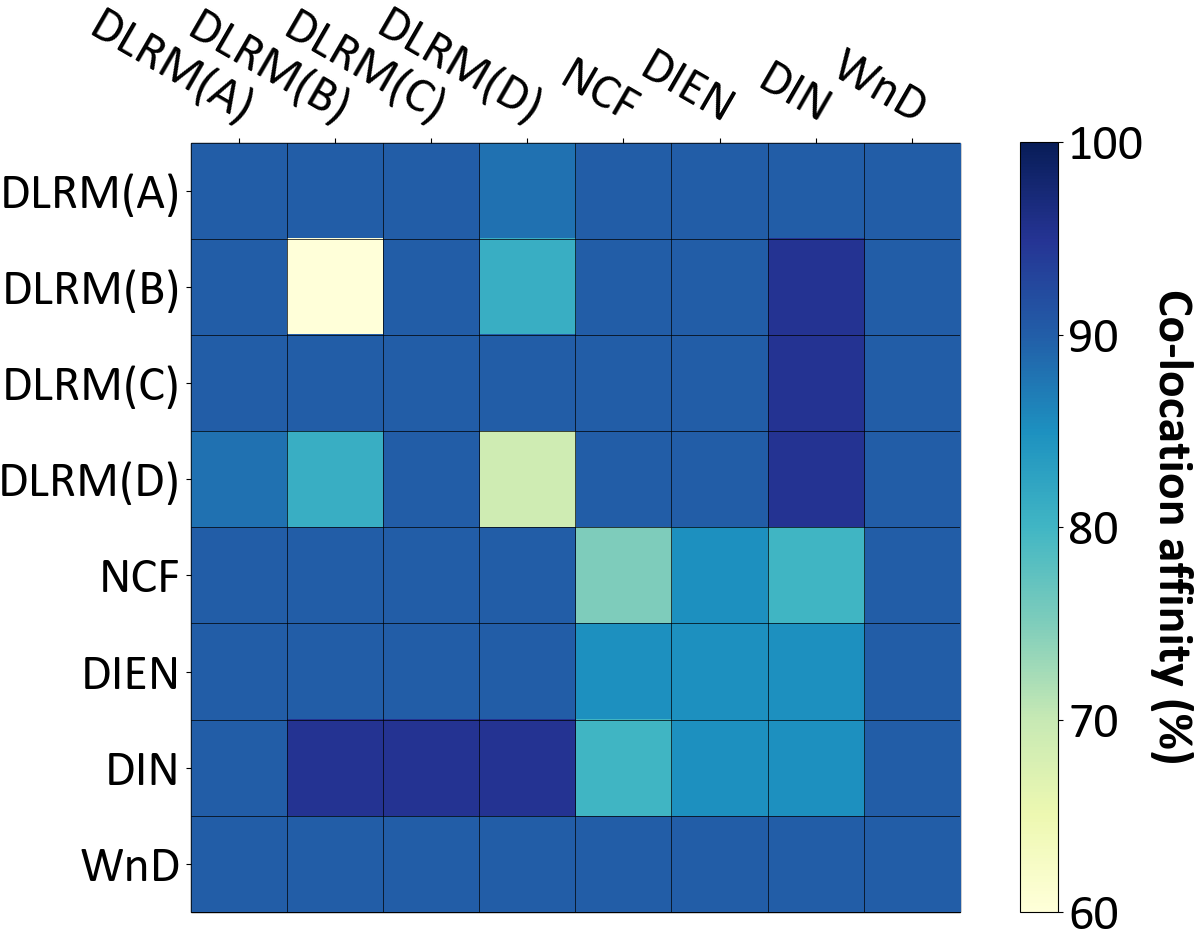}
}
\subfloat[] {
\includegraphics[width=0.2\textwidth]{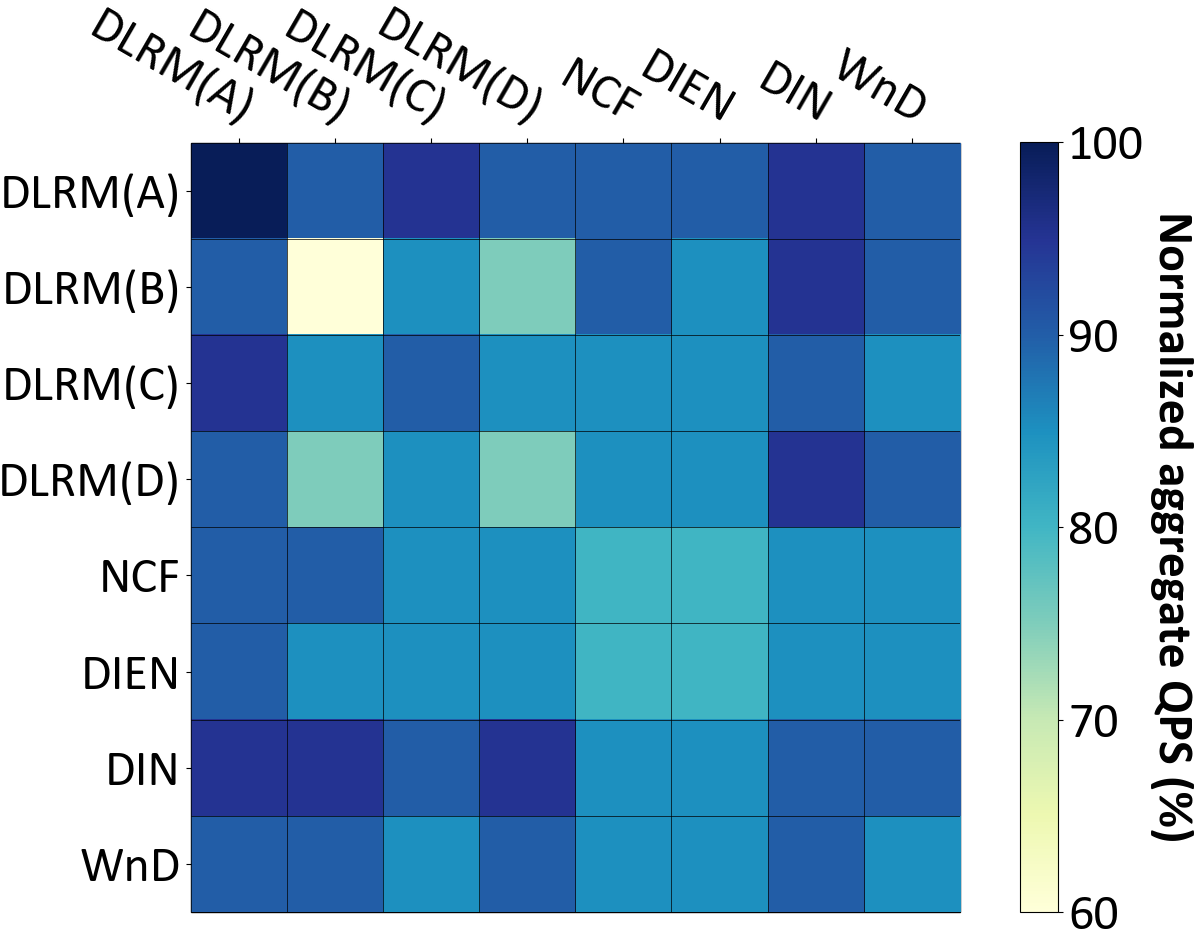}
}
\vspace{-0.5em}
\caption{
In (a), we show the \emph{estimated} co-location affinity among all possible
  pairs of co-located models (the higher the better). To demonstrate how well
  our co-location affinity models the effect of shared resource interference,
      we show in (b) the \emph{measured} aggregate QPS of co-located models
        normalized to the summation of the QPS achieved when each model is
        executed in isolation.  }
\vspace{-1.5em}
\label{fig:colocation_affinity}
\end{figure}

	Estimating co-location affinity for memory bandwidth sharing (\coaff{DRAM})
	by following the same measure as done in deriving \coaff{LLC} is challenging.
	This is because there is currently no practical way to manually partition and
	isolate each model's memory bandwidth usage (unlike the LLC where Intel's
			Cache Allocation Technology~\cite{intel_cat} provides means to fine-tune
			LLC partitioning). We therefore employ the analytical model in line $13$
	which utilizes our profiled result in
	\fig{fig:sensitivity_memory_hierarchy}(b) to measure co-location affinity
	with respect to memory bandwidth sharing. Here \membw{A} and \membw{B}
	are the amount of memory bandwidth consumed when each model is given half of
	the CPU cores and the entire LLC for isolated execution without co-location.
	By normalizing the sum of \membw{A} and \membw{B} to the available
	socket-level memory bandwidth (\membw{system}), we get an estimate on how
	much \emph{effective} bandwidth each model will be able to utilize
	vs. an idealistic scenario without bandwidth interference due to co-location
	(line $13$). The evaluated \coaff{DRAM} can therefore provide guidance on how
	intrusive the memory bandwidth sharing will have on co-located models,
	quantifying co-location affinity at memory.  For a conservative evaluation of
	co-location affinity that considers interference at	both LLC and memory, we
	choose the lower value among the \coaff{LLC} \coaff{DRAM} (line $16$).
	\fig{fig:colocation_affinity}(a) plots the derived co-location affinity for
	all possible model pairs we study. To visualize how well our 
	co-location affinity captures the interference and its effect on QPS, we also
	show in \fig{fig:colocation_affinity}(b) the \emph{measured} aggregate QPS of
	co-located models normalized to the theoretically maximum QPS achievable when
	each model is executed in isolation. As shown, 
	\fig{fig:colocation_affinity} clearly demonstrates the strong correlation
	between our estimated co-location affinity and the measured QPS (i.e., Pearson
			correlation coefficient: $0.95$)

\begin{algorithm}[t!]
\caption{\proposed Cluster Scheduling Algorithm}
\label{algo:cluster_scheduler}
\begin{algorithmic}[1]
\scriptsize
\rmfamily
	\State $\mathrm{TargetQPS[m_{1}, m_{2}, ..., m_{n}] = [QPSm_{1}, QPSm_{2}, ..., QPSm_{n}]}$
	\State $\mathrm{ServicedQPS[m_{1}, m_{2}, ..., m_{n}] = [0, 0, ..., 0]}$
	\State $\mathrm{Low     =  [models \  with  \ low  \ worker \  scalability]}$
	\State $\mathrm{High    = [models \  with \  high \  worker \ scalability]}$
\\
\LineComment{Step A: Allocate servers to be co-located with low worker scalability models}
	\For {$\mathrm{m_{i} \ \textbf{in} \ Low}$}
	\While {$\mathrm{ServicedQPS[m_{i}]  <  TargetQPS[m_{i}]}$}
		\State $\mathrm{m_{j}  =  find\_model\_with\_highest\_colocation\_affinity(m_{i},  High)}$
		\State $\mathrm{AllocateNewServer(m_{i}, m_{j})}$:
		\State $//~\mathrm{qps_{m_{i}}, qps_{m_{j}}:\ maximum \ QPS\ when \ m_{i} \ \& \ m_{j} \ are  \ co-located }$
		\State $\mathrm{ServicedQPS[m_{j}] \ += qps_{m_{i}}}$
		\State $\mathrm{ServicedQPS[m_{i}] \ += qps_{m_{j}}}$
	\EndWhile
\EndFor
\\
\LineComment{Step B: Allocate servers for high worker scalability models}
	\For {$\mathrm{m \ \textbf{in} \ High}$}
	\While {$\mathrm{ServicedQPS[m] < TargetQPS[m]}$}
		\State $\mathrm{AllocateNewServer(m)}$
		\State $//~\mathrm{qps_{m}: \ maximum \ QPS \ when \  m  \ is \  executed \ in \ isolation}$
		\State $\mathrm{ServicedQPS[m] \ += qps_{m}}$
	\EndWhile
\EndFor
\end{algorithmic}
\end{algorithm}

It is worth pointing out that the parameters to derive
\algo{algo:colocation_affinity} are all statically determined.  Therefore, the
derivation of co-location affinity for all possible model pairs are done
offline and are stored as a lookup table inside a two-dimensional array
(indexed using \model{A} and \model{B}'s identifier,
 \fig{fig:colocation_affinity}(a)). This table is then utilized by the central
master node with global cluster visibility to determine model pairs to
co-locate in inference servers to achieve a cluster-wide target QPS.
\algo{algo:cluster_scheduler} is a pseudo-code of \proposed's cluster
scheduling algorithm where we start by examining the low worker scalability
models for deployment, checking the co-location affinity table to find the best
candidate for co-location (line $6$).  Specifically, the scheduler prioritizes
models with high worker scalability as prime candidates for co-location with
low worker scalability models. When low worker scalability models are all
deployed, the remaining models are allocated with a dedicated server in
isolation but with maximum workers spawned to maximize QPS (line $17$).

\subsection{Node-level Resource Management Unit}
\label{sect:hera_resrc_manager}

Once the model pair to co-locate are chosen for an inference server, \proposed goes through the
server \emph{initialization} procedure for bootstrapping, followed by
an iterative \emph{monitor-and-adjust} process to periodically fine-tune the
resource allocation policy per inference query traffic (\fig{fig:hera_overview}).

{\bf Initialization.} Server bootstrapping is done by evenly partitioning the
CPU cores and shared LLC among the co-located workers (i.e., one core per each
		worker, all workers for a single model allocated with half the LLC),
		properly allocating the required data structures
		\emph{in-memory}. If one of the co-located model does not have enough
		worker scalability to fully utilize allocated cores (e.g., memory
				capacity limited \dlrm{B}), the \emph{other} model utilizes those idle
		cores to spawn more workers.

\begin{algorithm}[t!]
\caption{\proposed Resource Management Algorithm}
\label{algo:rmu}
\begin{algorithmic}[1]
 \scriptsize
\rmfamily
\Procedure {resource\_management\_routine()}{}
\While {True}
\LineComment{Step A: Monitor phase}
	\State $\mathrm{Monitor \ tail \ latancy, \ QPS, \ and \ query \ traffic \ rate \ for \ T_{monitor}}$
\LineComment{Step B: Adjust phase}
	\For {$\mathrm{\textbf{each} \ model \ M}$}
	\State $\mathrm{slack = tail\_latency_{M}/SLA_{M}}$
	\If {$\mathrm{slack > 1.0 \ or \ slack < 0.8}$}
	\State $\mathrm{adjust\_workers(M)}$
\EndIf
\EndFor
	\If {$\mathrm{number \ of \ workers \ have \ changed}$}
	\State $\mathrm{adjust\_LLC\_partition()}$
\EndIf
\EndWhile
\EndProcedure
\\
\Procedure {adjust\_workers}{$m$}
	\State $\mathrm{urgency = tail\_latency_{m}/SLA_{m}}$
	\If {$\mathrm{urgency < 1.0}$}
	\State $\mathrm{urgency = 1}$ 
\EndIf
	\State $\mathrm{adjusted\_traffic = urgency \times traffic_{query}}$
	\State $\mathrm{number\_of\_workers = find\_number\_of\_workers(m, adjusted\_traffic)}$
	\State $\mathrm{Allocate \ Model \ m \ with \ number\_of\_workers}$

\EndProcedure
\\
\Procedure {adjust\_LLC\_partition()}{}
	\State $\mathrm{QPS_{highest} = 0}$
	\For {$\mathrm{i = 1 \ \textbf{to} \ CacheWay_{max}}$}
	\State $\mathrm{CacheWay_{A} = i}$
	\State $\mathrm{CacheWay_{B} = CacheWay_{max} - CacheWay_{A}}$
	\State $\mathrm{QPS_{curr} =} \newline \mathrm{QPS[Model_{A}][Number \ of \ workers \ allocated \ for \ Model_{A}][CacheWay_{A}] +} \newline \mathrm{QPS[Model_{B}][Number \ of \ workers \ allocated \ for \ Model_{B}][CacheWay_{B}]}$
	\If {$QPS_{curr} \  > \  QPS_{curr}$}
		\State $\mathrm{QPS_{highest} = QPS_{curr}}$
		\State $\mathrm{best\_partition = (CacheWay_{A},\ CacheWay{B})}$
	\EndIf
\EndFor
	\State $\mathrm{Allocate \ Model_{A} \ and \ Model_{B} \ with \ best\_partition}$
\EndProcedure
\end{algorithmic}
\end{algorithm}

{\bf Monitor.} After server initialization, \proposed resource management unit
(RMU) periodically monitors tail latency, QPS, and the service query traffic
rate for a period of \texttt{T$_{monitor}$} to evaluate the effectiveness of
current resource allocations (line $4$ in \algo{algo:rmu}). The RMU checks
whether the current core/LLC allocation is appropriate or not by calculating
the SLA slack, i.e., ratio between measured latency vs. the model's SLA target
(line $7$).  If the tail latency is larger than the SLA, the RMU deems that the
current resource allocation is under-provisioned and the likelihood of
violating SLA is too high.  Contrarily, if the tail latency is smaller than the
SLA, the possibility of SLA violation is relatively low. However, too large of
an SLA slack by some predefined threshold ($80\%$ of SLA in our default
    setting) also implies that the current resource allocation is unnecessarily
over-provisioned.  Therefore, when either one of these conditions are met for a
given model (line $8$), RMU calls functions \texttt{adjust\_workers()} and
\texttt{adjust\_LLC\_partition()} to properly upsize/downsize the allocated
cores and LLC to make sure they are given sufficient amount of resources to
meet SLA (line $9$,$13$).

{\bf Adjusting model workers.} At-scale datacenters receive a dynamically
fluctuating query arrival patterns following a Poisson distribution
(\sect{sect:methodology}).  SLA violations occur because the rate at which
inference queries arrive to the server is overwhelmingly too high for the
currently available workers. Therefore, the ability to dynamically provision a
proportional amount of workers and LLC per query arrival rate is crucial for
appropriately handling query-level parallelism.  Now recall that our
characterization on worker scalability (the left-axis in
    \fig{fig:worker_scalability}) provides a proxy on how much sustained QPS
can be achieved with a given number of workers.  The RMU therefore utilizes the
performance scalability curve in \fig{fig:worker_scalability} (the same data
    structure used in \sect{sect:hera_model_selection} for deriving models'
    worker scalability) to find out the minimum number of workers that can
achieve sustainable QPS for the current input query arrival rate,
        \texttt{traffic$_{query}$} (\texttt{find\_number\_of\_workers()} in
            line $24$), the result of which is used to adjust the number of
        workers for the next monitoring phase. Given the reactive nature of
        \algo{algo:rmu} (i.e., allocating additional workers occur \emph{after}
            SLA violation is observed), we additionally define \emph{urgency}
        of a model's queries as defined in line $19$-$21$ to make sure
        \proposed can adequately handle high spikes in query arrival rates.  We
        define urgency as the ratio between the measured tail latency and the
        SLA target, so the higher the urgency the more likely that the
        inference server has delinquent service queries yet to be serviced
        inside the server request queue, leading to an unusually high tail
        latency. By artificially scaling up the observed query traffic rate
        with its urgency, the RMU helps better provision large enough workers
        for urgent models, enabling agile responsiveness. Of course, such
        over-provisioning might lead to unnecessarily large SLA slack after
        adjustment, but such case is gracefully handled by the RMU by properly
        downsizing the workers during the next monitor-and-adjust phase.

{\bf Adjusting LLC partitions.} When the number of workers for each co-located
model has changed, the RMU also adjusts the partitioned LLC ways as appropriate
using \texttt{adjust\_LLC\_partition()}.  Similar to how the optimal LLC way
partitioning design point in \algo{algo:colocation_affinity} was derived, we
employ a profiling-based strategy.  Specifically, a one-time, offline profiling
of QPS for all the models for all possible combination of (number of workers,
    number of LLC ways) is conducted, which is used to populate the
(three-dimensional) lookup table utilized in line $33$ of \algo{algo:rmu}.
Whenever \texttt{adjust\_LLC\_partition()} is called, the RMU utilizes this
lookup table to re-evaluate the optimal number of LLC ways to allocate under
the renewed (upsized/downsized) number of workers allocated for each model that
results in the highest aggregate QPS.  The overhead of generating this lookup
table is amortized over all future deployments over a target server
architecture and the memory allocation required to store this data structure is
less than $2$ KBs, having negligible impact on performance and memory usage.

\section{Evaluation} 
\label{sect:results}

Our evaluation takes a bottom-up approach, first focusing on
intra-node evaluation (\sect{sect:eval_constant_ld},
		\sect{sect:eval_diurnal_ld}), followed by our cluster-wide analysis
(\sect{sect:server_utility}). Following prior work~\cite{parties}, we first
evaluate scenarios where the multi-tenant workers run at constant loads
(\sect{sect:eval_constant_ld}) and later explore fluctuating load
(\sect{sect:eval_diurnal_ld}).

\subsection{Constant Load}
\label{sect:eval_constant_ld}

\subsubsection{Effectiveness of Hera Model Selection Unit}
\label{sect:eval_model_sel}

We establish two baseline model selection algorithms and two \proposed based
design points as follows.  The baseline designs are: 1) state-of-the-art
\texttt{DeepRecSys}~\cite{deeprecsys}, which co-locates multiple workers from a
single, \emph{homogeneous} model
(\sect{sect:related}/\sect{sect:effect_multitenant}), and 2) \emph{randomly}
choosing any given pair of \emph{heterogeneous} models to co-locate without any
restriction (\random).  The two \proposed design points are 3) \randomplus and
4) \hera, where both utilize the worker scalability of the candidate model to
avoid the undesirable co-location of (high, high) worker scalability models.
However, \randomplus randomly chooses any one of the possible model pairs
(excluding (high, high) model pairs), whereas \hera can utilize our estimated
co-location affinity (\fig{fig:colocation_affinity}) to choose model pairs that
provide the \emph{highest} machine utilization.  To rule out the effect of
resource management policy in our evaluation, all four design points including
\proposed employ our proposed resource management algorithm in
\sect{sect:hera_resrc_manager}.

We measure \emph{Effective Machine Utilization} (EMU), a metric used in related
prior work~\cite{heracles,parties,clite} that is defined as the max aggregate
load of all co-located applications, where each application's load is expressed
as a percentage of its \emph{max load} when executed in isolation with all
server resources (\fig{fig:colocation_example}, \sect{sect:effect_multitenant}
    discusses how max load is measured for each model).  Note that EMU can be
above $100\%$ by better bin-packing shared resources among co-located models.

\begin{figure}[t!] \centering
\includegraphics[width=0.42\textwidth]{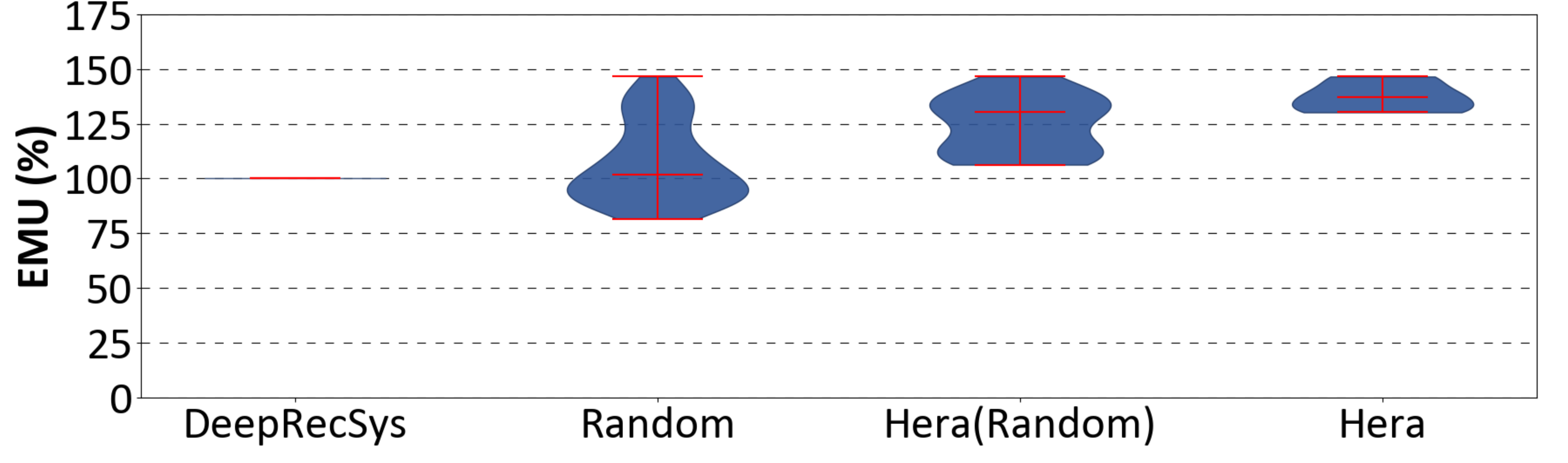}
\caption{
  Violin plots of EMU with constant load. The three markers represent min,
         median, and max EMU.
}
\vspace{-1.5em}
\label{fig:emu}
\end{figure}

\fig{fig:emu} shows the distribution of EMU for each model selection
algorithm's all chosen pair of co-located models.  Because \deeprecsys does not
co-locate workers from multiple models, QPS is always identical to the max load
of isolated execution, thus having EMU of $100\%$.  As for \random, we show the
EMU distribution for all possible combination of model pairs among the eight
models we study, achieving $82$ to $147\%$ EMU. Although \random can improve
EMU when opportunistically co-locating low worker scalability models with other
models, it fails to rule out the co-location of (high, high) scalability
co-location pairs with low co-location affinity, resulting in worst-case $18\%$
EMU loss. 

The two \proposed design points on the other hand utilizes worker scalability
to successfully rule out model pairs with low co-location affinity,
   guaranteeing the EMU never falls below $100\%$ and achieve substantial EMU
   improvement than \deeprecsys and \random.  However, a key difference between
   \randomplus and \hera are the following. When the cluster-level scheduler
   selects model pairs to co-locate, \randomplus makes random selection from
   any model combinations except (high, high) worker scalability model pairs.
   In contrast, \hera utilizes the estimated co-location affinity to
   judiciously choose model pairs with the highest EMU, leading to significant
   EMU improvements vs. the other three data points.  Overall, \hera achieves
   $37.3\%$, $34.7\%$, and $5.4\%$ average EMU improvement than \deeprecsys,
   \random, and \randomplus, respectively. 

\begin{figure}[t!] 
\centering
\subfloat[PARTIES]{
\includegraphics[width=0.20\textwidth]{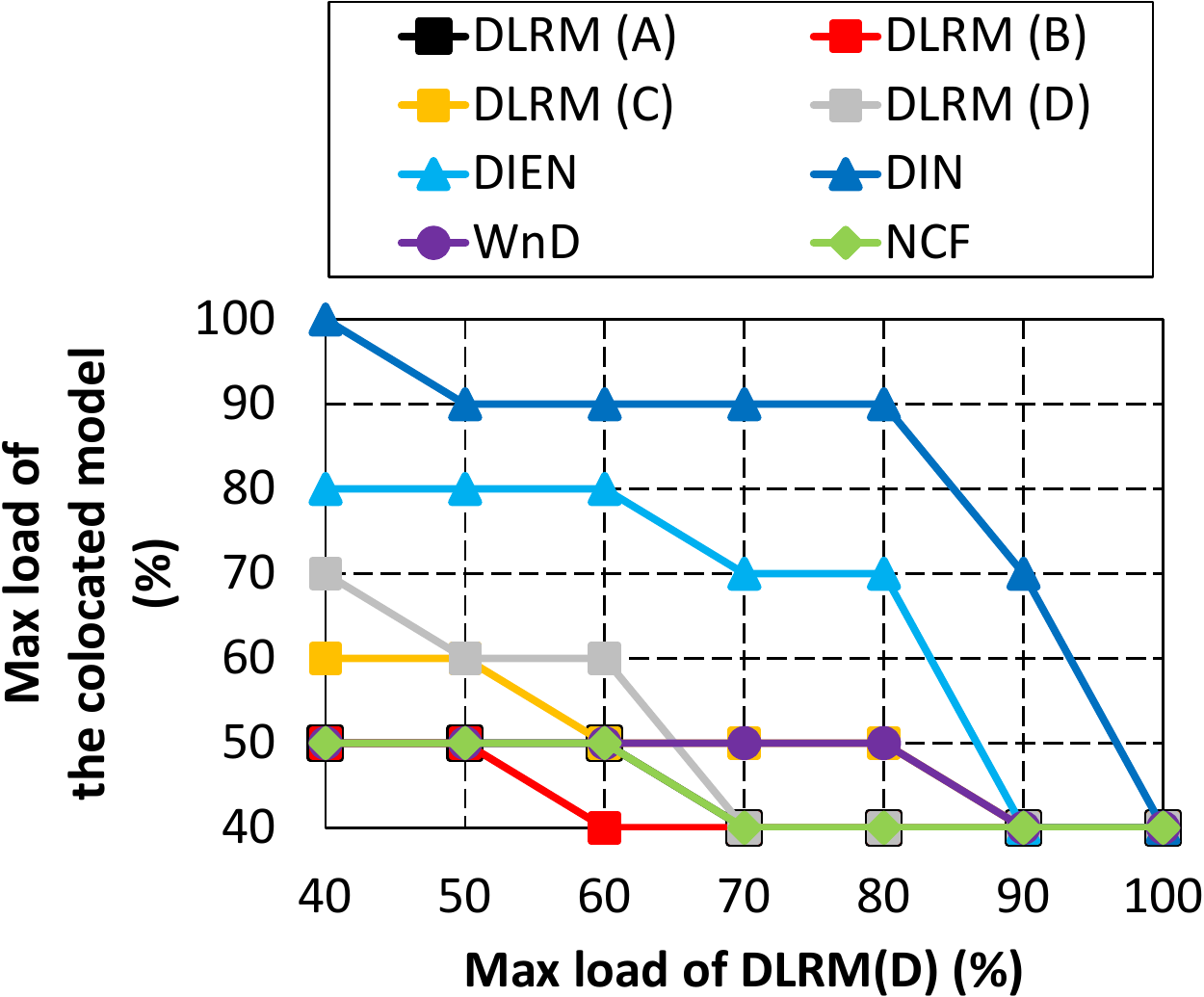}
	\label{fig:dlrm_d_parties}
}
\subfloat[Hera]{
	\includegraphics[width=0.165\textwidth]{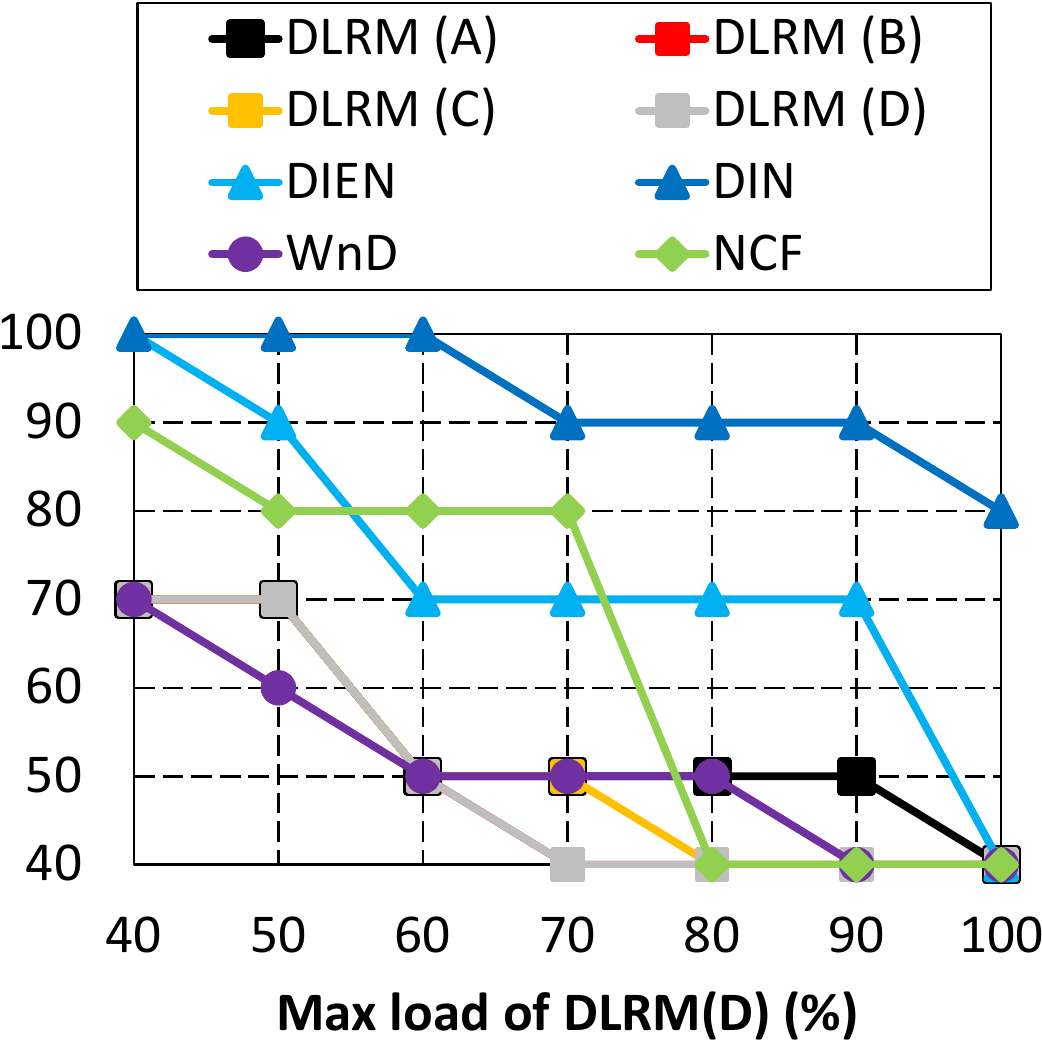}
	\label{fig:dlrm_d_hera}
}
\caption{ 
Each plot shows the result of co-locating the memory bandwidth limited \dlrm{D}
(Model$_{x}$) with each of our studied recommendation models (Model$_{y}$)
  using (a) PARTIES and (b) \proposed. Each line shows the maximum percentage
  of Model$_{y}$'s \emph{max load} (y-axis) that can be achieved without SLA
  violation when Model$_{x}$ (i.e., \dlrm{D}) is running at the fraction of its
  own \emph{max load} indicated on the x-axis. Therefore, the aggregation of
  max load for Model$_{x}$ and Model$_{y}$ at a particular design point equals
  its EMU. For instance, the EMU of \dlrm{D} and \ncf when \dlrm{D} is at
  $50\%$ max load is ($50$$+$$50$)$=$$100\%$ under PARTIES, while \proposed
  achieves ($50$$+$$80$)$=$$130\%$.
}   
\vspace{0.5em}
\label{fig:max_load_dlrm_d}
\end{figure}

\subsubsection{Effectiveness of Hera Resource Manager}
\label{sect:eval_resrc_manager}

\begin{figure}[t!] 
\centering
\subfloat[Co-location with \ncf]{
\includegraphics[width=0.20\textwidth]{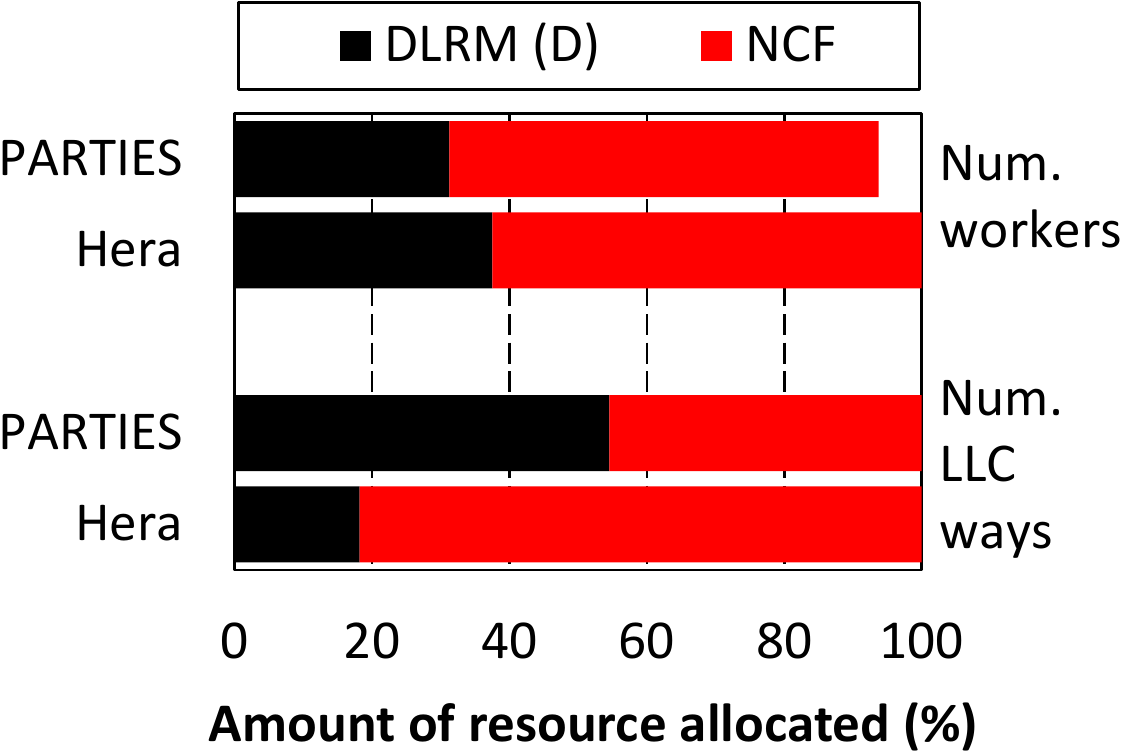}
	\label{fig:latency_resnet}
}
\subfloat[Co-location with \din]{
	\includegraphics[width=0.20\textwidth]{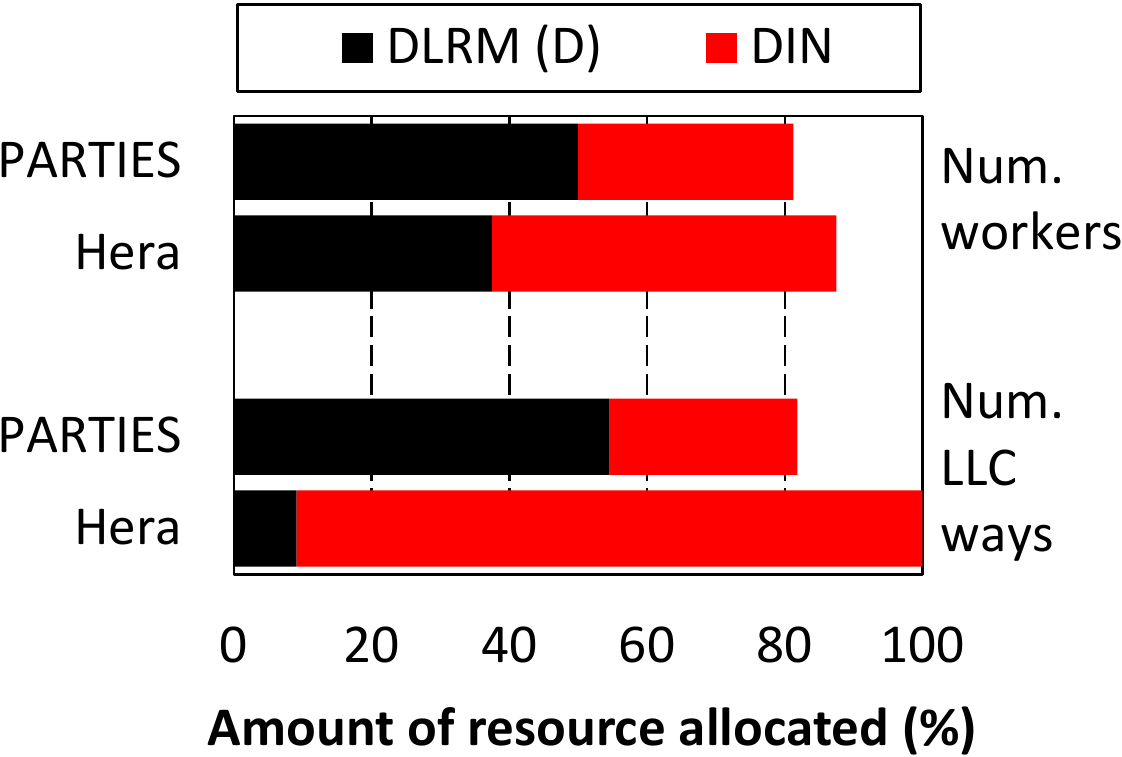}
	\label{fig:latency_gnmt}
}
\caption{ 
Number of workers and LLC ways allocated when the low worker scalability
  \dlrm{D} with $50\%$ max load is co-located with high worker scalability (a)
  \ncf and (b) \din. \proposed successfully allocates sufficient number of
  workers and LLC ways, reaching $80\%$/$100\%$ of max load for \ncf/\din.
  PARTIES fails to provision high enough LLC to cache-sensitive \ncf/\din, only
  achieving $50\%$/$90\%$ of max load for these two models.
}   
\vspace{-1em}
\label{fig:dlrm_d_snapshot}
\end{figure}

PARTIES~\cite{parties} is a QoS-aware intra-node resource manager targeting
generic, latency-critical cloud services. To clearly demonstrate the novelty of
\proposed's application-aware resource manager, we implement PARTIES on top of
\proposed's model selection algorithm and compare its EMU against \proposed.
Across all evaluated scenarios, \proposed achieves an average $12\%$ (maximum
    $55\%$) EMU improvement than PARTIES.  Due to space constraints, we show in
\fig{fig:max_load_dlrm_d} a subset of our evaluation where we visualize the max
load of low worker scalability \dlrm{D} (x-axis) and the other co-located model
(y-axis).  Hera generally achieves higher sustained max load over PARTIES for
the majority of deign points when gradually injecting each model with loads
from $40\%$ to $100\%$ of their respective max load.  The reason behind
\proposed's superior performance is twofold. First, \proposed is based on our
profile-based characterization to determine a good initial starting point
within the search space to find the optimal core/LLC resource allocation
scheme. Second, while PARTIES needs to carefully fine-tune all shared resources
in a system (i.e., in addition to core/LLC, the disk and network bandwidth is
    also carefully monitored and adjusted by PARTIES), our application-aware
\proposed leverages the unique properties of ML inference servers (e.g.,
    in-memory processing, \sect{sect:hera_model_selection}) to narrow down the
resource management targets, thus enabling rapid determination of an optimal
resource management strategy that best suits recommendation model's
heterogeneous memory needs. In \fig{fig:dlrm_d_snapshot}, we show a snapshot of
PARTIES vs.  \proposed's resource allocation when the low worker scalability
\dlrm{D} is co-located with high worker scalability \ncf and \din. Given its
cache sensitive property (\fig{fig:cache_way_sensitivity}), \ncf and \din
require sufficient LLC capacity to sustain high max load. While PARTIES is able
to eventually allocate enough workers for \ncf, the amount of LLC provisioned
for this cache-sensitive workload is much lower than under \proposed, failing
to achieve high max load.

\subsection{Fluctuating Load}
\label{sect:eval_diurnal_ld}

\begin{figure}[t!] \centering
\includegraphics[width=0.44\textwidth]{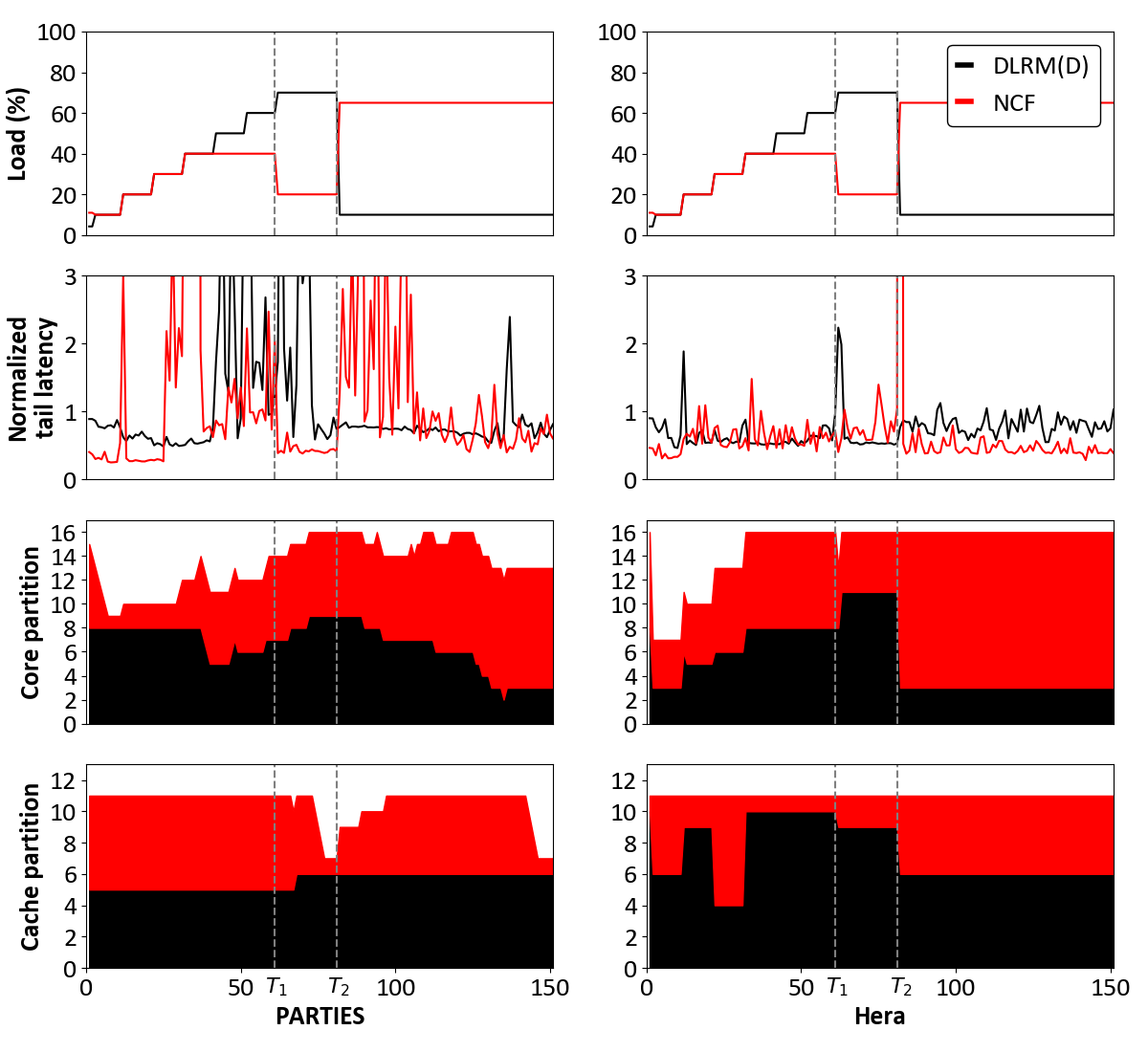}
\caption{
Changes in tail latency and resource allocations with \proposed and PARTIES
  with fluctuating load for \dlrm{D} and \ncf. Tail latency is normalized to
  each model's respective SLA targets, i.e., normalized latency larger than
  $1.0$ signifies a QoS violation. Note that the maximum number of cores and
  LLC ways is $16$ and $11$ under our studied evaluation
  (\tab{tab:server_config}).
}
\vspace{-1.5em}
\label{fig:dynamic_parties_vs_hera}
\end{figure}

This section evaluates the robustness of \proposed's resource manager vs.
PARTIES in handling dynamically changing query arrival rates.  At-scale
datacenter applications frequently experience fluctuations in their load (e.g.,
    sudden burst of high traffic load, diurnal patterns where load is high at
    daytime while gradually decreasing during night time).  To simulate such
scenario, we employ measures proposed by Chen et al.~\cite{parties} where we
co-locate models with heterogeneous resource requirements, e.g., \dlrm{D} and
\ncf, and vary the query arrival rate as illustrated in
\fig{fig:dynamic_parties_vs_hera}: the load to both \dlrm{D} and \ncf are
gradually increased until \loadtime{1}, which is when \ncf experiences a sudden
decrease in its load. At \loadtime{2}, the query arrival rate to \ncf is
suddenly spiked from $20\%$$\rightarrow$$60\%$ of its max load while \dlrm{D}
sees a sudden drop from $70\%$$\rightarrow$$10\%$.
\fig{fig:dynamic_parties_vs_hera} clearly shows that \proposed does a much
better job in maintaining tail latency at below SLA, unlike PARTIES which
frequently exhibits sudden spikes of SLA violations throughout its execution.
Our analysis revealed that PARTIES, while it \emph{is} capable of eventually
figuring out a decent resource allocation strategy, the decision is based on a
constant upsize/downsize feedback loop that monitors various shared resources
within the system.  Because \proposed utilizes the profile-based
characterization lookup table (\algo{algo:rmu}), it is able to rapidly
determine the optimal resource allocation even at sudden changes of loads at
\loadtime{1} and \loadtime{2}, enabling robust execution even under fluctuating
query arrival rates.

\subsection{Cluster-wide Server Utilization}
\label{sect:server_utility}

In this section, we analyze \proposed's effectiveness in reducing the number of
servers required to fulfill a \emph{cluster-wide} target QPS goal.  Because the
set of models deployed by hyperscalers as well as each model's expected service
demand (i.e., query arrival rate) are proprietary information not publicly
available, we take the following measure for our evaluation.

\begin{figure}[t!] \centering
\includegraphics[width=0.44\textwidth]{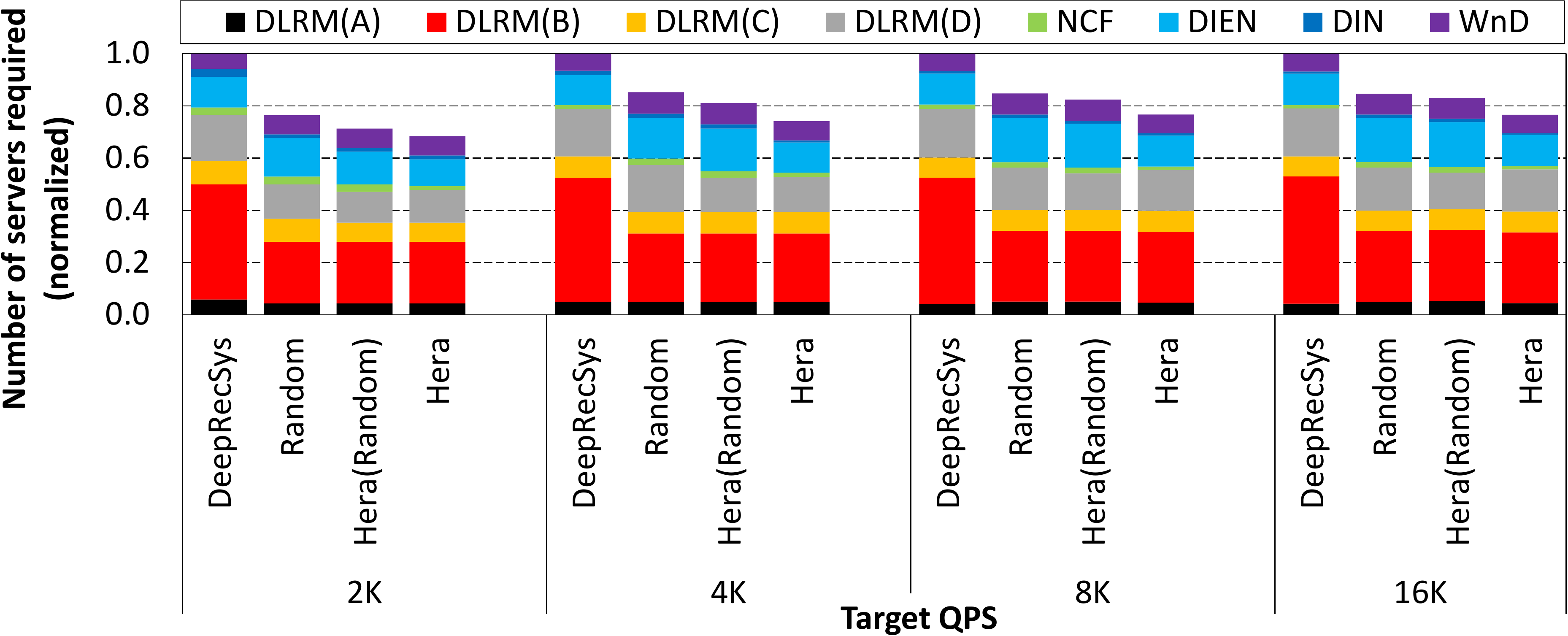}
\caption{
The number of servers required (y-axis) to service a target QPS level (x-axis) when
the target QPS per each model is evenly distributed in an identical manner.
}
\label{fig:num_of_servers_required}
\end{figure}

{\bf Even distribution of target QPS among models.} We first assume that the
eight models in \tab{tab:model_config} have an \emph{identical} level of target
QPS (\fig{fig:num_of_servers_required}).  We then utilize the four
model selection algorithms  to measure the total number of servers required to
satisfy this target QPS.  When the aggregate target QPS across all the models
exceeds the max load serviceable by our multi-node cluster, we run separate
rounds of experiments in an iterative manner to quantify how many additional
servers are needed to service the remaining QPS. 

The baseline \deeprecsys allocates a single server to service a single model,
    so low worker scalability models like \dlrm{B,D} require a larger number of
    servers to reach the target QPS. As for \random, the cluster scheduler
    randomly selects a given pair of models for multi-tenant deployment until
		the target QPS is achieved. Because \random is capable of better utilizing
		compute nodes servicing low worker scalability models like \dlrm{B,D}, it
    is able to reduce the number of servers by an average $15\%$ compared to
    \deeprecsys. However, \random is not able to properly isolate the
    interference at shared resources, so the unproductive co-location of (high,
        high) worker scalability models suffers from low efficiency.  Both
    \randomplus and \hera are worker scalability aware and can potentially
    avoid deploying model pairs with low co-location affinity (e.g., \ncf vs.
        \dien/\din/\wnd). Nonetheless, \hera does a much better job reducing
    the number of servers vs. \randomplus as it effectively utilizes our
    estimated co-location	affinity to minimize unfruitful model co-location
    pairs.  For instance, \hera noticeably reduces the number of servers
    utilized for deploying \dien all	 thanks to its ability to better bin-pack
    the shared resources among co-located models, achieving an average $26\%$
    and  $11\%$ reduction in total
required servers than \deeprecsys and \random, respectively. 

\begin{figure}[t!] \centering
\includegraphics[width=0.44\textwidth]{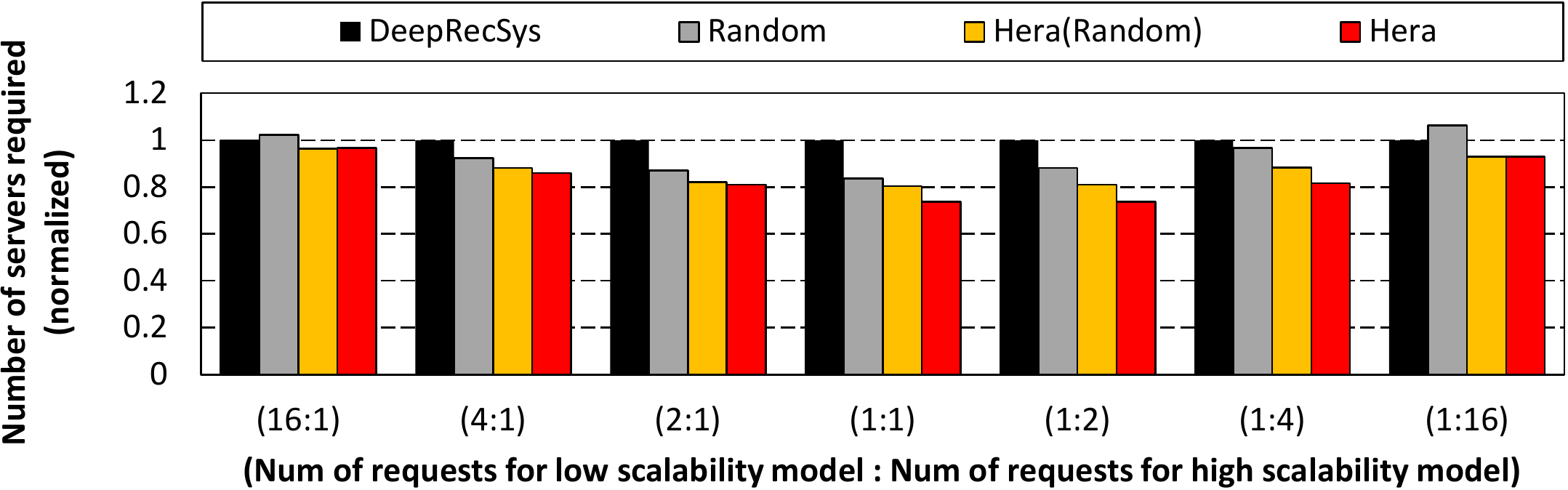}
	\caption{
The number of servers required (y-axis) when low and high worker scalability models
have different ratios of target QPS levels (x-axis).
}
\label{fig:num_of_servers_required_popularity}
\end{figure}

{\bf Skewed distribution of target QPS among models.}
\fig{fig:num_of_servers_required_popularity} shows the number of servers needed
when the per-model target QPS exhibits a skewed distribution.  Unless the
queries are \emph{all} requested to low worker scalability models or all to
high scalability models (which is unlikely the common case in real-world
    settings), \hera is able to noticeably reduce the number of servers
required to reach a target QPS.

\subsection{Sensitivity}
\label{sect:hera_sensitivity}

\begin{figure}[t!] 
\centering
\vspace{-1.2em}
\subfloat[]{
\includegraphics[width=0.215\textwidth]{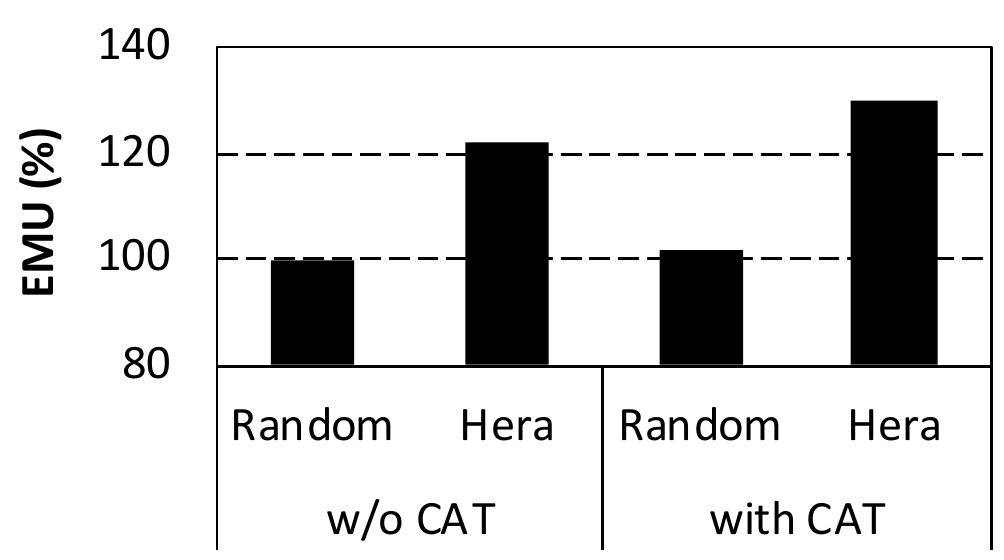}
	\label{fig:sensitivity_cat}
}
\subfloat[]{
	\includegraphics[width=0.215\textwidth]{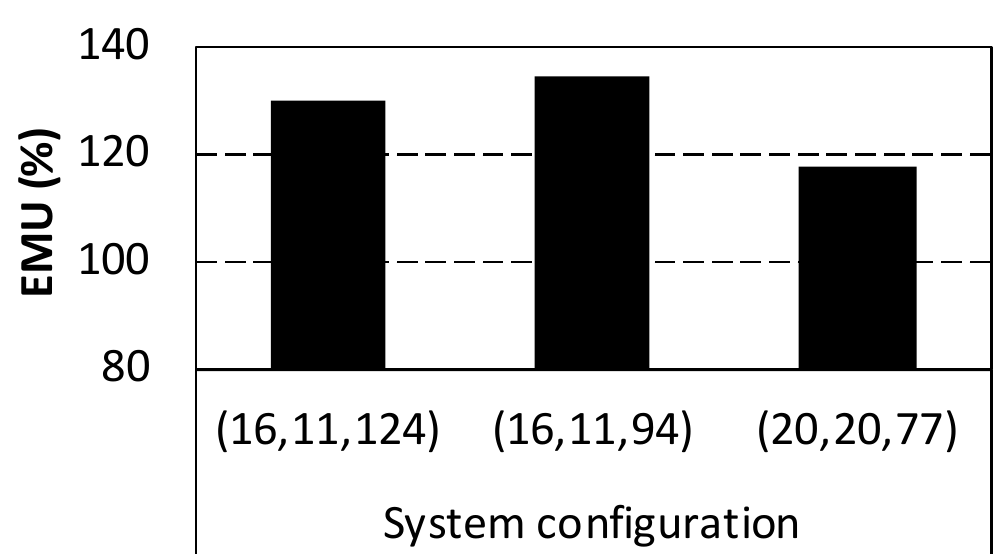}
	\label{fig:sensitivity_system}
}
\vspace{-0.25em} 
\caption{ 
Hera sensitivity to (a) enabling/disabling LLC partitioning (with and w/o CAT),
     and (b) different system configurations (the numbers inside the bracket
         refers to (number of cores, number of LLC ways, and memory bandwidth
           (GB/sec))). Note that the scale of the y-axis is $80$$-$$140\%$.
}   
\vspace{-0.5em}
\label{fig:sensitivity}
\end{figure}

{\bf Effect of Hera co-location and LLC partitioning.} Hera consists of two key
components: 1) affinity-aware co-location algorithm, and 2) application-aware
LLC partitioning via Intel CAT. To clearly quantify where Hera's main benefits
come from, we quantify the impact of isolating our two main proposals in
\fig{fig:sensitivity}(a) as an ablation study.  Even without LLC partitioning,
Hera's co-location algorithm alone provides $22\%$ EMU improvements vs.
Baseline DeepRecSys, with further $8\%$ improvement with LLC partitioning.

{\bf Different system configuration.} 
We show a subset of our sensitivity study to different system configurations in
\fig{fig:sensitivity}(b), illustrating the EMU improvement when the
underlying system configuration is changed with different number of CPU cores,
LLC ways, and memory bandwidth (GB/sec). As depicted, the
benefits of Hera remains intact across diverse system platform configurations,
achieving $30\%$/$35\%$/$18\%$ EMU improvement for the shown configurations.

\subsection{Design Overhead}
\label{sect:design_overhead} 

{\bf Profiling/deployment cost.} Hera is implemented as a profiling-based,
feedback-driven scheduler, and the main design overhead comes from profiling
the following two lookup tables: (a) QPS as a function of number of parallel
workers (\fig{fig:worker_scalability}) and (b) QPS as a function of LLC ways
allocated (\fig{fig:cache_way_sensitivity}).  The profiling time taken to
generate these two tables ($T_{worker}$ and $T_{LLC}$, respectively) are:

$\mathrm{T_{worker} = O(number \ of \ CPU \ cores)}$

$\mathrm{T_{LLC} = O(number \ of \ LLC \ ways \times number \ of  \ CPU cores)}$

Using our baseline system (\tab{tab:server_config}),
$T_{worker}$ and $T_{LLC}$ takes less than $1$ minute and $15$ minutes per each
model.  As each data point can be collected completely independently, profiling
and generating \fig{fig:worker_scalability} and \fig{fig:cache_way_sensitivity}
across hundreds of models can easily be done within tens of minutes assuming
thousands of compute nodes are available (which is readily accessible in cloud
datacenters). Utilizing these two lookup tables as a 2D software array,
generating the co-location affinity matrix in \fig{fig:colocation_affinity}(a)
(\algo{algo:colocation_affinity}) even for hundreds of models takes less than
one second using a single CPU core.

{\bf Deploying Hera across thousands of servers.} Each inference server is
deployed individually per-node and the execution of Hera's cluster-wide
scheduling algorithm (\algo{algo:cluster_scheduler}) incurs less than $100$ ms
of latency, enabling scalable deployment in at-scale datacenters.

\section{Conclusion}
\label{sect:conclusion}

\proposed is an application-aware co-location algorithm and resource management
software for multi-tenant recommendation inference. We first conduct a 
 characterization  on multi-tenant recommendations, uncovering its
heterogeneous memory capacity and bandwidth demands.  We then utilize such
property to develop a profiling-based, feedback driven \proposed runtime system
that dynamically adapts resource allocation among multi-tenant workers for
balancing latency throughput.  Compared to state-of-the-art,
\proposed significantly improves effective
machine utility while guaranteeing SLA.

\bibliographystyle{ieeetr}
\bibliography{references}

\end{document}